\documentclass[twocolumn,prx,nofootinbib,superscriptaddress]{revtex4-2}
\usepackage[utf8]{inputenc}
\usepackage[T1]{fontenc}
\usepackage{amsmath}
\usepackage{hyperref}
\usepackage{epsfig}
\usepackage{amssymb}
\usepackage{physics}
\usepackage{tikz}
\usetikzlibrary{quantikz2}
\usetikzlibrary{fit, backgrounds}
\usepackage{bm}

\usepackage{enumitem}

\usepackage[toc,page]{appendix}
\usepackage[caption=false]{subfig}
\usepackage{amsthm}

\hypersetup{colorlinks=true, linkcolor=blue, citecolor=blue, urlcolor=blue, unicode=true}
\usepackage{cleveref}
\Crefname{equation}{Eq.}{Eqs.}

\newcommand{\R}{\mathbb{R}}

\newcommand{\btheta}{{\bm{\theta}}}

\newcommand{\bphi}{\bm{\phi}}

\newcommand{\Ad}{\mathrm{Ad}}
\newcommand{\ad}{\mathrm{ad}}
\newcommand{\Span}{\mathrm{span}}

\newcommand{\SU}{\mathrm{SU}}
\newcommand{\U}{\mathrm{U}}

\newcommand{\SO}{\mathrm{SO}}
\newcommand{\Sp}{\mathrm{Sp}}
\newcommand{\su}{\mathfrak{su}}
\newcommand{\so}{\mathfrak{so}}

\newcommand{\zz}{\mathfrak{z}}
\renewcommand{\sp}{\mathfrak{sp}}

\newcommand{\g}{\mathfrak{g}}
\newcommand{\uu}{\mathfrak{u}}
\newcommand{\kk}{\mathfrak{k}}
\newcommand{\mm}{\mathfrak{m}}

\newcommand{\hh}{\mathfrak{h}}

\usepackage{xcolor}
\usepackage{adjustbox}
\definecolor{MyRed}{rgb}{0.49, 0.63, 0.98}
\definecolor{MyBlue}{rgb}{1,0.56,0.53}
\definecolor{MyBackground}{RGB}{128,255,185}
\newcommand{\myBrick}{%
\begin{quantikz}[column sep=2mm]
    &\gate[2]{}  \gategroup[6,steps=3, style={rounded
        corners, fill=MyBackground!60, inner xsep=5pt}, background]{} & &\gate[1, style={dashed}]{} \wire[d][5]{q}& \\
    & &\gate[2]{} & &\\
    &\gate[2]{} & & &\\
    & &\gate[2]{} & &\\
    &\gate[2]{} & & &\\
    & & & \gate[1, style={dashed}]{} & 
\end{quantikz}
}

\begin{document}

\title{Geometric Quantum Machine Learning with Horizontal Quantum Gates}
\author{Roeland Wiersema}
\affiliation{Vector Institute, MaRS  Centre,  Toronto,  Ontario,  M5G  1M1,  Canada}
\affiliation{Department of Physics and Astronomy, University of Waterloo, Ontario, N2L 3G1, Canada}
\affiliation{Xanadu, Toronto, ON, M5G 2C8, Canada}

\author{Alexander F. Kemper}
\affiliation{Department of Physics, North Carolina State University, Raleigh, North Carolina 27695, USA}

\author{Bojko N. Bakalov}
\affiliation{Department of Mathematics, North Carolina State University, Raleigh, North Carolina 27695, USA}

\author{Nathan Killoran}
\affiliation{Xanadu, Toronto, ON, M5G 2C8, Canada}

\date{May 30, 2024}
\begin{abstract}
    In the current framework of Geometric Quantum Machine Learning, the canonical method for constructing a variational ansatz that respects the symmetry of some group action is by forcing the circuit to be equivariant, i.e., to commute with the action of the group. This can, however, be an overzealous constraint that greatly limits the expressivity of the circuit, especially in the case of continuous symmetries.
    We propose an alternative paradigm for the symmetry-informed construction of variational quantum circuits, based on homogeneous spaces, relaxing the overly stringent requirement of equivariance.
    We achieve this by introducing horizontal quantum gates, which only transform the state with respect to the directions orthogonal to those of the symmetry. 
    We show that horizontal quantum gates are much more expressive than equivariant gates, and thus can solve problems that equivariant circuits cannot. For instance, a circuit comprised of horizontal gates can find the ground state of an $\SU(2)$-symmetric model where the ground state spin sector is unknown--a task where equivariant circuits fall short.
    Moreover, for a particular subclass of horizontal gates based on symmetric spaces, we can obtain efficient circuit decompositions for our gates through the KAK theorem. Finally, we highlight a particular class of horizontal quantum gates that behave similarly to general $\SU(4)$ gates, while achieving a quadratic reduction in the number of parameters for a generic problem. 
\end{abstract}
\nopagebreak
\maketitle
\section{Introduction\label{sec:introduction}}

Symmetries play a critical role in many scientific endeavours. 
In physics, if we can identify the symmetries of a system, then we can incorporate them into the equations we use to describe the system. This can help us simplify and solve otherwise intractable problems, identify important physical quantities (quantum numbers, conserved currents, invariants), and deepen our understanding of the system under study.
In machine learning, symmetries have also proven extremely powerful~\cite{bronstein2021geometric}. Understanding the underlying geometric regularities of real-world data, and adapting machine learning models to account for it, enables us to overcome the dreaded curse of dimensionality. The most powerful modern machine learning models, such as convolutional networks, graph neural networks, and transformers, all have deep symmetry underpinnings.

It is evident that quantum computing and quantum machine learning (QML) may also benefit greatly when we can identify and leverage symmetries. In recent years, the first forays into a study of \emph{Geometric QML} have begun~\cite{zheng2023geqml,meyer2023gqml, sauvage2024building, larocca2022gqml}. Notably, a standard recipe has emerged for ``geometrizing'' a QML model (specifically, a quantum circuit) under a known symmetry group.
This process, called \emph{twirling}, converts every gate (or layer) in the circuit into a new version that now has the property of \emph{equivariance} under the symmetry~\cite{seki2020symmadapted,ragone2022representation,sauvage2024building}. Combined with a final measurement which is invariant under the same symmetry, the modified circuit is now guaranteed to respect the symmetry, i.e., it gives the same output for all inputs differing only by a symmetry transformation.

Equivariance—essentially, the property where a gate commutes with group transformations—provides a natural mathematical condition for enforcing symmetries. It also presents a clear-cut recipe, based on twirling, for how to incorporate those symmetries into a circuit. However, the approach based on equivariant gates has a number of limitations. Practically, the twirling procedure can be expensive, potentially involving an exponentially large sum (for discrete symmetry groups, e.g., the symmetric group), or a difficult integral (for continuous symmetry groups). The procedure can also sometimes lead to entire classes of gates being dropped from the circuit, limiting expressivity. In fact, as we will demonstrate, requiring equivariance is too strong a condition; while it does lead to a notable reduction in the dimension of the circuit manifold, and guarantees that symmetries are recognized, it can often cause legitimate symmetry-respecting evolutions to be completely removed from consideration.

In our zeal to symmetrize a quantum circuit, we may thus find ourselves inadvertently underparameterizing it, leading to poorer performance. The drawbacks of obeying symmetries too strictly have been explored in a deep learning context, where it has been argued that \emph{approximate} equivariance may benefit training geometric neural networks~\cite{elsayed2020revisiting, wang2022approxeml}. 
With this in mind, we can ask if there exist less restrictive---but still symmetry-respecting---methodologies we can draw on for the quantum case to achieve the goals of Geometric QML. 

In this work, building upon the theoretical insights of~\cite{wierichs2023symm}, we identify a new approach for incorporating symmetries in quantum circuits, while providing greater freedom than strict equivariance. We propose a new class of symmetry-respecting gates, called \emph{horizontal} quantum gates. These gates have the notable property that they only rotate in directions on the unitary gate's manifold that are perpendicular to the directions generated by a symmetry. Intuitively, they can be thought of as gates that act on symmetry-equivalence classes of states rather than on individual state vectors. Horizontal gates can be parameterized and used in a variational quantum circuit, or used with fixed parameter values in any other quantum algorithm where symmetries are important. We construct these gates from mathematical objects called \emph{homogeneous spaces}, which naturally arise when one quotients out a (symmetry) subgroup from a larger Lie group. These spaces also arise commonly in the context of Geometric Deep Learning~\cite{cohen2018spherical}, the geometrization of quantum mechanics~\cite{doebner1975quantum,provost1980riemannian}, optimal control~\cite{dalessandro2021}, and have recently been studied in the context of barren plateaus~\cite{arvind2023quantum}.

We will explore the usage of horizontal quantum gates in a variety of different settings. In \Cref{sec:background}, we define the underlying mathematical ideas and describe the construction of horizontal gates from homogeneous spaces. In \Cref{sec:horizontal}, we highlight the limits of equivariant gates and show how horizontal quantum gates get around these issues. Then, in \Cref{sec:symmetric} we discuss a particular type of horizontal quantum gate constructed from a symmetric space (a special type of homogeneous space), which comes with a recipe allowing a decomposition into lower-dimensional quantum gates. Finally, in \Cref{sec:stabilizing} we discuss the possibility of constructing horizontal gates by looking at the stabilizer of a set of states, which leads to powerful gates that outperform general $\SU(4)$ operations. As a guiding light, we revisit and discuss the Bloch sphere in each section, since it is the simplest non-trivial example of a homogeneous space.

\section{Background\label{sec:background}}
\subsection{Quantum Gates}
A variational quantum circuit corresponds to a product of parameterized unitaries $U(\btheta_{1:L}) = \prod_{l=1}^L U_l(\btheta_l)$ acting on an $n_q$-qubit Hilbert space $\mathcal{H} = (\mathbb{C}^2)^{\otimes n_q}$, where $\btheta_l\in\mathbb{R}^{d_l}$ and $\btheta_{1:L} =\{\btheta_1,\ldots,\btheta_L\}$. 
A single quantum gate $U(\btheta)$ acting on $n\leq n_q$ qubits can be parameterized as follows:
\begin{align}
    U(\btheta) &= \exp{A(\btheta)},\quad A(\btheta) = \sum_{j=1}^{d} \theta_j H_j, \label{eq:gate}
\end{align}
where $\theta_j\in\mathbb{R}$, $\{H_j\}$ is a set of skew-Hermitian operators of size $N\times N$, and $N=2^n$. Recall that a vector space that is closed under the commutator is a Lie algebra, and we can generate a Lie algebra from $\{H_j\}$ by taking nested commutators. The resulting Lie algebra is called the \emph{dynamical Lie algebra} $\g$ (see~\cite{Albertini2001dynlie, albertini2021subspace,chen2017preparing,wang2016subspace,dalessandro2021, wiersema2023dla}). We can always think of $A(\btheta)$ as an element of $\g$ and $U(\btheta)$ as an element of the corresponding Lie group $G = e^\g$. In this work, we will only consider the case where $\g$ is a compact semi-simple Lie algebra, so that the Lie group $G = e^\g$ is compact.
In general, it is known that $\g$ is always \emph{reductive}, so it is a direct sum of a compact semi-simple Lie algebra and a center commuting with all of $\g$ (see, e.g., Proposition A.1 in \cite{wiersema2023dla} for a proof).

The dynamical Lie algebra of a quantum gate is intricately related to the directions that can be explored on the full unitary group in a gradient-based optimization. In particular, the gradient of a variational quantum gate is of the form
\begin{align}
    \partial_{\theta_j} U(\btheta) = \underbrace{U(\btheta)}_{\in G}\underbrace{\Omega_j(\btheta)}_{\in\g} \in T_{U(\btheta)} G, \label{eq:gradient}
\end{align}
where $T_{U(\btheta)}G$ denotes the tangent space of $G$ at the point $U(\btheta)$ and $\Omega_j(\btheta)$ is called the \emph{effective generator}. This object can be calculated classically, as long as the gate acts locally, i.e., its matrix representation is low-dimensional (see~\cite{wiersema2023here, kottmann2023evaluating}). It is also possible to classically compute the effective generator by using the adjoint representation, in cases where the associated dynamical Lie algebra is small \cite{heidari2024efficient}.

\subsection{Homogeneous Spaces\label{sec:hom}}
Let $G$ again be a compact semi-simple Lie group with a unitary representation on the complex Hilbert space $\mathcal{H}$, so that every $g\in G$ is represented with a unitary matrix. The action of $G$ defines an equivalence relation of states 
\begin{align*}
    g\cdot \ket{\psi} \sim \ket{\psi},\quad\forall g\in G, \; \ket{\psi}\in\mathcal{H}.
\end{align*}
We denote by $[\psi]$ the equivalence class  of $\ket{\psi}$, i.e., its orbit under the action of $G$. A physically inspired example is the case where $G=\U(1)$, i.e., quantum states that differ by a global phase are equivalent because they are indistinguishable under the Born rule.

Consider a symmetry represented by a Lie group $K$, where $K$ is a closed subgroup of $G$. We can use the action of $K$ on $G$ to define an equivalence relation on $G$:
\begin{align*}
    g \sim g \cdot k,\quad \forall k \in K.
\end{align*}
The set of all elements equivalent to $g$ is called a (left) \emph{coset} $gK$ of $K$ in $G$. 
These cosets can be used to construct a smooth manifold
\begin{align*}
    G/K = \{gK \,|\, g \in G\}.
\end{align*}
The space $G/K$ is a \emph{homogeneous space}, which means that any two elements in $G/K$ can be transformed into each other via the action of some $g\in G$ (see \cite{arvanitogeorgos2003introduction} for a review). Given that $G$ is compact, we also know that $G/K$ is reductive, hence $\g$ can be decomposed into two subspaces,
\begin{align}
    \g = \kk\oplus\mm \label{eq:decomp},
\end{align}
where $\kk$ is the Lie algebra of $K$. For compact $G$ as considered here, $\g$ has a negative-definite symmetric invariant bilinear form, the trace inner product, and
we can take $\mm =\kk^\perp$ to be the orthogonal complement of $\kk$. 
Note that $\kk$ is a subalgebra of $\g$, i.e., it is closed under the commutator and is itself a Lie algebra. On the other hand,
while $\mm$ is not a Lie algebra in general, it instead has the important property that $\Ad(k) \mm \subset \mm$ for all $k\in K$, where $\Ad(k)(X) = k X k^{-1}$ for $X\in\mm$~\cite{arvanitogeorgos2003introduction}. It can then be concluded that
\begin{align}
    [\kk,\kk] \subseteq \kk,\quad [\kk,\mm] \subseteq \mm. \label{eq:homogeneous}
\end{align}
This means that the subspace $\mm$ is invariant under symmetry transformations.

\begin{figure}[htb!]
    \centering
    \includegraphics[width=0.45\textwidth]{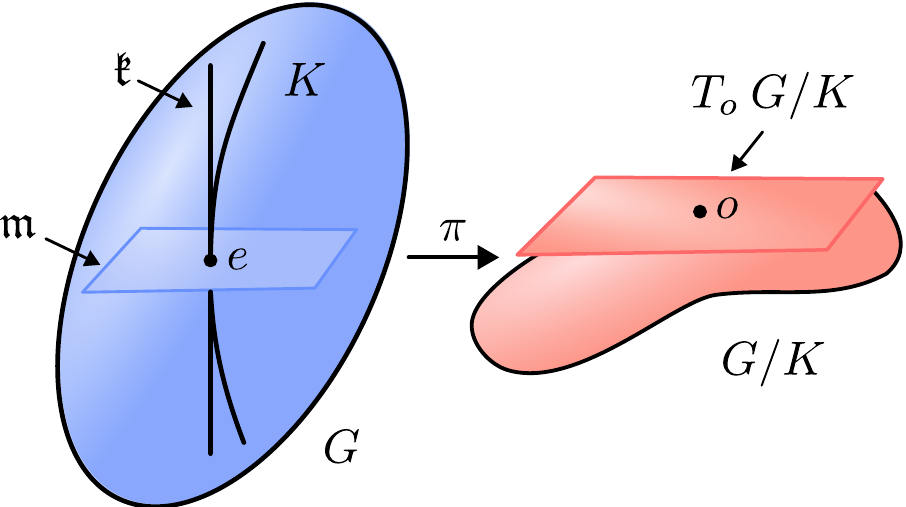}
    \caption{\textbf{Schematic representation of the tangent space of a homogeneous space.} The coset $eK$ gets taken to a single point $o$ in the homogeneous space $G/K$ via a projection $\pi$. The projection leaves only the horizontal space $\mm$ as the tangent space of $G/K$. Figure adapted from~\cite{oneill1983semiriemannian}.}
    \label{fig:homspace}
\end{figure}
We will call $\kk$ the \emph{vertical} subspace and $\mm$ the \emph{horizontal} subspace of $\g$ with respect to $K$. The horizontal directions in $\mm$ correspond to $T_o(G/K)$, which is the tangent space to the homogeneous space $G/K$ at the point $o=eK$ (see App. \ref{app:tang_hom}). We illustrate this correspondence schematically in \Cref{fig:homspace}.

As discussed in~\cite{wierichs2023symm}, the split in~\Cref{eq:decomp} can be further decomposed by introducing the notion of an \emph{equivariant} subspace $\g^{\kk}$, which is defined as the commutant (or centralizer) of $\g$ with respect to $\kk$:
\begin{align}
    \g^\kk = \{X \in \g \,|\, [X,Y]=0,\: \forall Y\in\kk\}. \label{eq:commutant}
\end{align}
We can decompose $\kk$ by splitting off its center
\begin{align}
    \zz(\kk) = \{X\in\kk\,|\, [X, Y] = 0,\:\forall Y \in \kk\}, \label{eq:center}
\end{align}
which gives $\kk=\zz(\kk)\oplus \kk_o$ where $\kk_o$ is the orthogonal complement to $\zz(\kk)$ within $\kk$. Note that $\g^\kk \cap \kk = \zz(\kk)\subseteq \zz(\g^\kk)$.
We then decompose $\g^\kk$ into
$\zz(\kk)$ and the orthogonal complement of $\zz(\kk)$ within $\g^\kk$, denoted by $\g^\kk_o$. This finally gives the decomposition
\begin{align}
    \g =\lefteqn{\underbrace{\phantom{\mathfrak{r} \oplus\g^\kk_o}}_{\mm}}\mathfrak{r} \oplus
    \lefteqn{\overbrace{\phantom{\g^\kk_o\oplus\zz(\kk)}}^{\g^\kk }}\g^\kk_o \oplus
    \lefteqn{\underbrace{\zz(\kk) \oplus \kk_o}_{\kk}.} \label{eq:wierichs_decomp}
\end{align}

\subsection{Equivariant Quantum Gates}
Let $K$ be a closed subgroup of $G$ with Lie algebras $\kk$ and $\g$, respectively. To parameterize all of $\g$, we choose the generators $\{H_j\}$ in \Cref{eq:gate} as a basis for $\g$. This choice ensures that $A(\btheta)\in\g$ for all $\btheta$. 
We now act with the symmetry element $k\in K$ on a quantum state $\ket{\psi}$, and subsequently act with the quantum gate from \Cref{eq:gate} on the resulting state:
\begin{align}
    U(\btheta) (k \cdot \ket{\psi}) = k\cdot U'(\btheta) \ket{\psi},\label{eq:commute_through}
\end{align}
where we have defined $U'(\btheta) = \exp{k^{-1} A(\btheta) k}$. Hence, in general, pulling the symmetry through the action of the gate requires a completely new gate $U'(\btheta)$ generated by a new generator $k^{-1} A(\btheta) k\in \g$. An equivalent statement is that 
\begin{align}
    [\kk, A(\btheta)] \subseteq \g,\label{eq:in_gg}
\end{align}
since $\g$ is closed under the commutator.
Note that in \Cref{eq:in_gg} the symmetry does not necessarily commute with the gate. This is the result of our choice of generators $\{H_j\}$, which does not use any information about $K$ in the gate construction.

On the other hand, equivariant quantum gates are parameterized so that, by design,
\begin{align}
    [\kk, A(\btheta)] = 0 \label{eq:equiv_cond}
\end{align}
for all $\btheta$, so $k^{-1} A(\btheta) k=A(\btheta)$, and $U(\btheta)$ commutes with the symmetry,
\begin{align*}
    U(\btheta) (k \cdot \ket{\psi}) = k\cdot U(\btheta) \ket{\psi}.
\end{align*}
From \Cref{sec:hom} it is clear how we can construct such a gate: we parameterize $A(\btheta)$ with the equivariant directions in the commutant (see \Cref{eq:commutant}),
\begin{align}
    U_{\mathcal{E}}(\btheta) = \exp{A_{\mathcal{E}}(\btheta)},\quad A_{\mathcal{E}}(\btheta) = \sum_{j=1}^{d} \theta_j E_j,\label{eq:U_eq}
\end{align}
where $E_j \in \g^\kk$. 
A powerful property of these gates is that composing them into a larger variational circuit will lead to an equivariant quantum circuit, which can be useful for learning functions that are invariant under the symmetry $K$. 

\begin{figure}[htb!]
    \centering
    \subfloat[Equivariant]{\includegraphics[width=0.45\columnwidth]{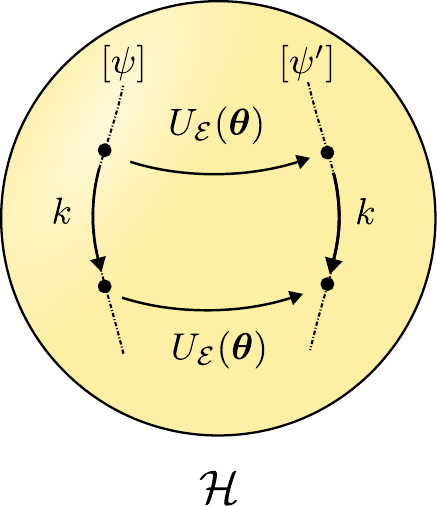}}
    \hspace{5mm}
    \subfloat[Horizontal]{\includegraphics[width=0.45\columnwidth]{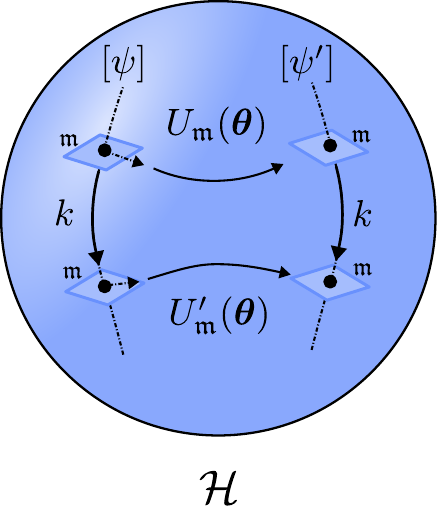}}
    \caption{\textbf{Equivariant versus horizontal quantum gates.} (a) We can move states around in the Hilbert space by acting with the symmetry $k$ and the equivariant circuit $U_\mathcal{E}(
    \btheta)$, and the order does not impact the result. (b) The action of $k$ can change the tangents of a horizontal quantum gate, but keeps them within the subspace $\mm$. Changing the order of transformations requires a change in the unitary $U_\mm(\theta)$.}
    \label{fig:eq_vs_hor}
\end{figure}
\subsection{Horizontal Quantum Gates}
A large drawback of equivariant quantum gates is that for many continuous symmetries, the commutant $\g^\kk$ is trivial (see \Cref{tab:dimensions}). Currently there is no way to account for such degenerate symmetries in the construction of variational quantum gates.
Here, we propose a method to take the symmetry $K$ into account by way of \emph{horizontal} quantum gates. These gates are generated by the horizontal directions $\mm$,
\begin{align}
    U_{\mm}(\btheta) = \exp{A_{\mm}(\btheta)},\quad A_{\mm}(\btheta) = \sum_{j=1}^{d} \theta_j M_j\label{eq:U_mm},
\end{align}
with $M_j \in \mm$. Note that by \Cref{eq:homogeneous}, we have the property
\begin{align}
    [\kk, A_{\mm}(\btheta)] \subseteq \mm. \label{eq:comm_Amm}
\end{align}
Hence the family $A_\mm(\btheta)$ always stays within $\mm$ when it is acted upon with a symmetry operation $k$, and does not move into the larger Lie algebra $\g$. Since $\exp{t X}$ for $X\in\mm$ is a geodesic on $G/K$, we can think about $U_{\mm}(\btheta)$ as the shortest path between two cosets $gK$ and $g'K$.
A horizontal quantum gate can thus be understood as a middle ground between the general condition of \Cref{eq:in_gg} and the overly strong constraint of equivariance in \Cref{eq:equiv_cond}. Gates which satisfy \Cref{eq:U_mm} are more expressive than their equivariant counterparts, while still taking the symmetry into account. 
We schematically display the difference between equivariant and horizontal quantum gates in \Cref{fig:eq_vs_hor}. 

We now have all the ingredients to understand how one can use horizontal gates 
for problems with continuous symmetries. Revisiting \Cref{eq:commute_through} in the case of horizontal gates $U_\mm(\btheta)=\exp{A_{\mm}(\btheta)}$, we see that 
\begin{align}
    \exp{A_{\mm}(\btheta)} (k \cdot \ket{\psi}) 
    &=  k\cdot \exp{k^{-1}A_{\mm}(\btheta)k} \ket{\psi} \nonumber\\
    &=  k\cdot \exp{A_{\mm}(\btheta')} \ket{\psi}.\label{eq:commute_through_horiz}
\end{align}
Since $k^{-1}A_{\mm}(\btheta)k\in\mm$ for horizontal gates, we determine that $\btheta'$ is simply a reparametrization of the generator defined within the horizontal subspace $\mm$ (\Cref{eq:U_mm}). 
Taking together multiple horizontal gates will result in a circuit where one can commute any symmetry action $k$ to the end, at the cost of reparametrizations. 
We emphasize that, unlike an equivariant circuit, a circuit consisting of horizontal gates can generate symmetry transformations in $K$. This occurs because, although $\mm$ is orthogonal to $\kk$, the subspace $\mm$ is not a Lie algebra, so hence the resulting unitary $\exp{A_{\mm}(\btheta)}$ can move the state in the direction of $K$. Therefore, the true benefit of horizontal gates is to avoid parametrizing the directions of the symmetry explicitly rather than never moving the state in that direction.



In the following sections, we will explore three different methods for constructing horizontal quantum gates. First off, if both $G$ and $K$ are given it is straightforward to construct $G/K$ and the corresponding horizontal gates. Secondly, we consider symmetric spaces, which are homogeneous spaces with additional properties that make them well suitable to decompose into simpler gates. Finally, we can consider states or sets of states and ask which operations leave them unaffected, which results in a third possible construction of a horizontal gate.

As a basis for $\su(N)$, we will use tensor products of Pauli matrices. For finding $\mm$, we use a variety of methods. Either we find it analytically, use \texttt{scipy.linalg.null\_space} to determine $\kk^\perp$ numerically, or take the involutions of App.\ \ref{sec:classification} to split off $\kk$ from $\g$ and project it onto a basis of $\g$. We provide code to reproduce the examples in this work at~\cite{our_code} and discuss the techniques to find the decomposition of \Cref{eq:wierichs_decomp} in App.~\ref{app:decomp}. For clarity, we use the notation ``$i\,\Span$'' to indicate that each element in the linear span is multiplied with the imaginary unit $i$.
Gradients of variational quantum circuits are calculated from \Cref{eq:gradient} via the method of~\cite{wiersema2023here, kottmann2023evaluating}, which allows one to calculate gradients of quantum gates generated by Hamiltonians with multivariate parameters. We do not consider shot noise, and perform all numerical simulations with PennyLane~\cite{bergholm2018pennylane}.

\begin{table}[htb!]
    \centering
    \resizebox{\columnwidth}{!}{
    \begin{tabular}{|c|c|c|c|c|c|c|}\hline
         $G$ & $K$ & 
         $\dim(\mathfrak{r})$ &  
         $\dim(\g_o^\kk)$ & 
         $\dim(\zz(\kk))$ & 
         $\dim(\kk_o)$ & Section\\\hline
         $\SU(2)$ & $\U(1)$ & 2 & 0 & 1 & 0 & \ref{sec:horizontal}-\ref{sec:stabilizing} \\
         $\SU(4)$ & $\SU(2)^{S=1/2}$& 11 & 1 & 0 & 3 & \ref{sec:horizontal}\\
         $\SU(4)$ & $\SU(2)\times\SU(2)$ & 9 & 0 & 0 & 6 & \ref{sec:symmetric}\\
         $\SU(4)$ & $\U(3)$ & 5 & 1 & 0 & 9 & \ref{sec:stabilizing}\\
         $\SO(4)$ & $\SO(3)$ & 3 & 0 & 0 & 3 & \ref{sec:stabilizing}\\
         $\SU(4)$ & $\SU(2)^{S=3/2}$ & 12 & 0 & 0 & 3 & App.~\ref{app:more}\\
         $\SU(4)$ & $\Sp(2)$ & 5 & 0 & 0 & 10 & App.~\ref{app:more}\\
         $\SO(4)$ & $\SU(2)$ & 3 & 0 & 0 & 3 & App.~\ref{app:more}\\
         $\SO(4)$ & $1\times \SO(2)\times 1$& 4 & 1 & 1 & 0 & App.~\ref{app:more}\\
         $\SU(8)$ & $\mathrm{S}(\U(2)\times \U(6))$ & 24 & 0 & 0 & 39 & App.~\ref{app:more}\\
         \hline
    \end{tabular}
    }
    \caption{\textbf{Homogeneous spaces considered in this work.} Note that $\g^\kk_o$ and $\zz(\kk)$ are zero for many of these spaces, which implies that there are no equivariant directions. }
    \label{tab:dimensions}
\end{table}
\subsection{From Gates to Quantum Circuits}

Since we work mostly with quantum gates, we need to address how local symmetries acting on one- or two-qubit systems translate to larger systems. We always consider $K$ to act locally and so its global action is given via a tensor power representation. 

An example would be the case where we consider a representation of the symmetry $\SU(2)$ acting on single-qubit subsystems (the spin-$1/2$ representation). The global tensor product representation of this symmetry would be
\begin{align}
    K_{\mathrm{circuit}} = \left\{k^{\otimes n_q} \,\big|\, k\in \SU(2)^{S=1/2}\right\}.
\end{align}
A two-qubit horizontal gate would then be an element of $\SU(4)/(\SU(2)\times\SU(2))$.
On the other hand, we could consider $\SU(2)$ acting on a 4-dimensional subsystem (the spin-$3/2$ representation). We then find
\begin{align}
    K_{\mathrm{circuit}} = \left\{k^{\otimes n_q/2} \,\big|\, k\in \SU(2)^{S=3/2}\right\}.
\end{align}
In this case, a two-qubit horizontal gate would then be an element of $\SU(4)/\SU(2)$. A quantum circuit composed of these individual gates would then take symmetry $\SU(2)^{S=3/2}$ into account on the entire space. 

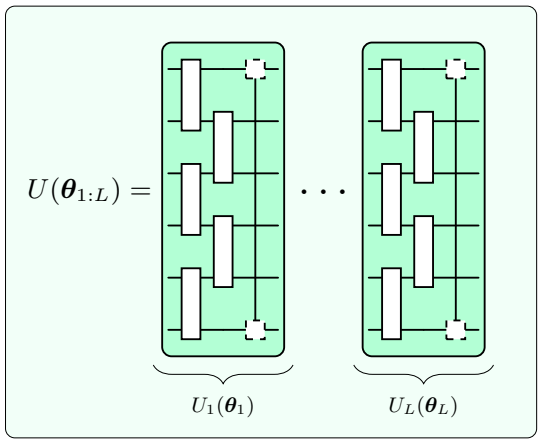
\begin{figure}[htb!]
\centering
\adjustbox{width=0.4\textwidth,valign=B}{
\begin{tikzpicture}
    \node[scale=1.3] (A0) at (-2,0) {$U(\btheta_{1:L})=$};
    \node (A1) at (0,0) {\myBrick};
    \node[scale=2] (A2) at (1.5,0) {$\hdots$};
    \node (A3) at (3,0) {\myBrick};
    \node (A4) at (4.4,-3.4) {};
    \draw[decorate,decoration={brace,amplitude=8pt,mirror}] (A1.south west) -- (A1.south east) node[midway,below=10pt] {$U_1(\btheta_1)$};
    \draw[decorate,decoration={brace,amplitude=8pt,mirror}] (A3.south west) -- (A3.south east) node[midway,below=10pt] {$U_L(\btheta_L)$};
    \begin{scope}[on background layer]
    \node[fit=(A0) (A1) (A2) (A3) (A4), draw, inner sep=5pt, rounded
    corners, fill=MyBackground!10] {}; 
    \end{scope}
\end{tikzpicture}}
 \caption{\textbf{Bricklayer variational quantum circuit}. The dashed gate at the edges of each layer indicate potential open or closed boundary conditions. We initialize each block with the identity.}
\label{fig:brick}
\end{figure}

\begin{figure*}[htb!]
    \centering
    \subfloat[Uniform]{ \includegraphics[width=0.5\textwidth]{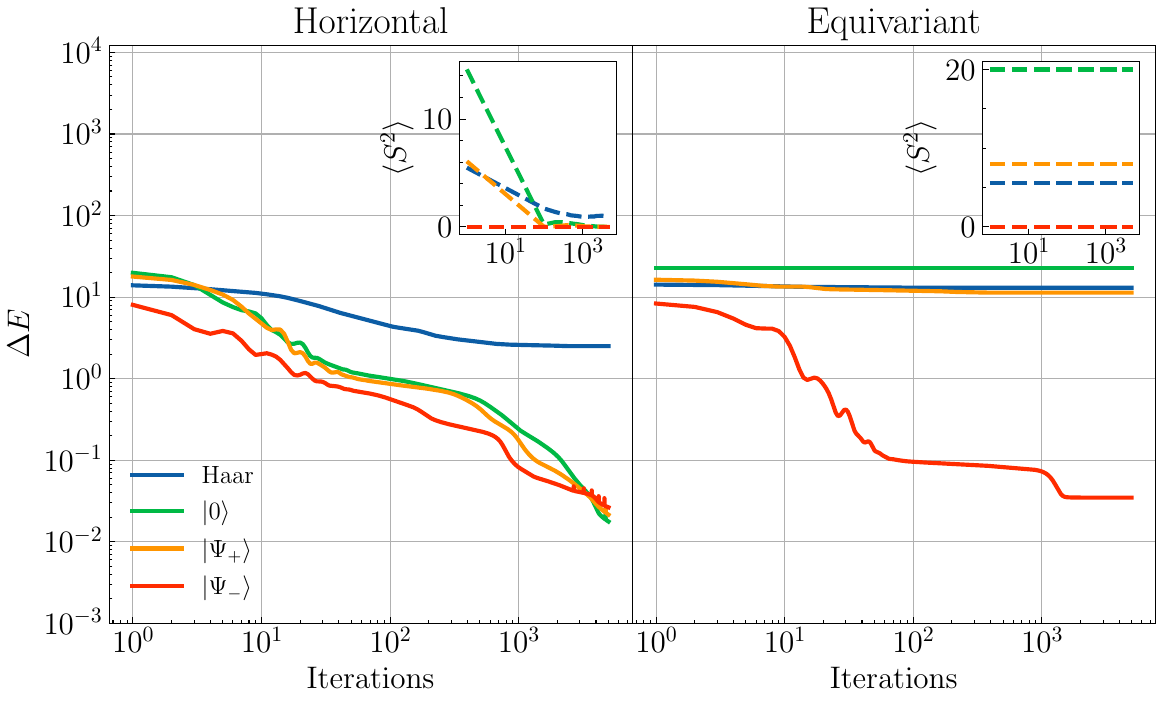}}
    \subfloat[Random]{ \includegraphics[width=0.5\textwidth]{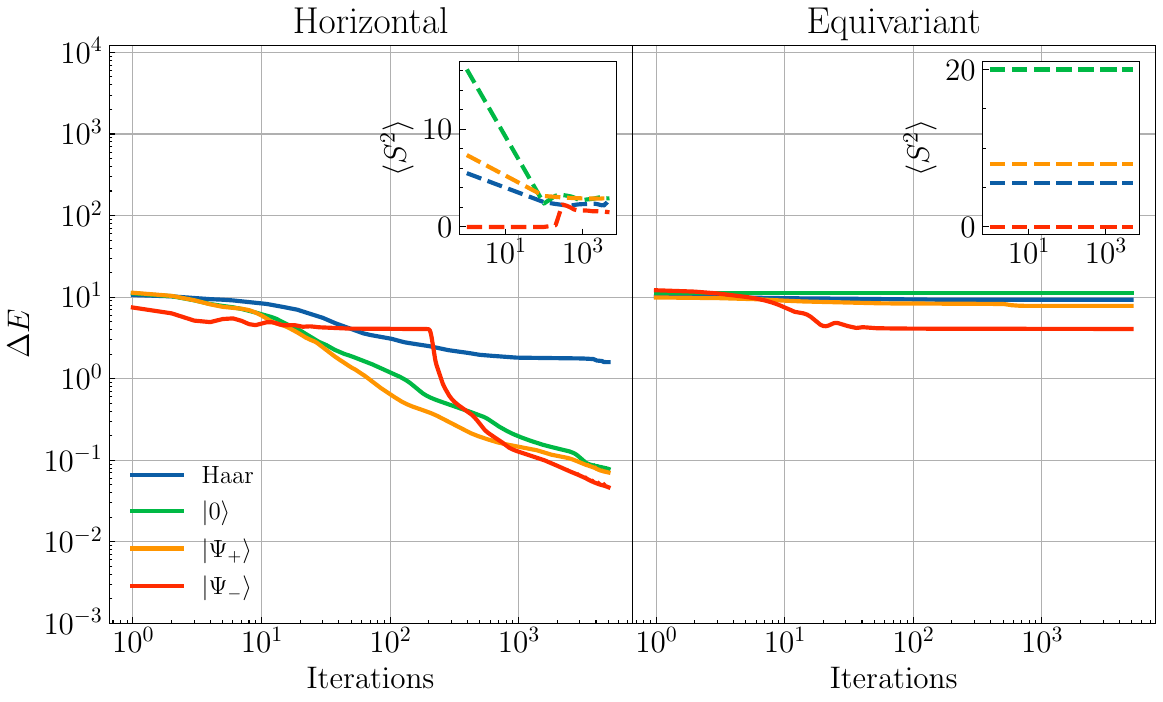}}
    \caption{{\textbf{Optimization trajectories for VQE on a Heisenberg chain of 8 qubits with a depth $8$ variational quantum circuit.}} We show the energy difference $\Delta {E} = E_{\mathrm{circuit}} - E_0$ at each iteration of the VQE optimization, where $E_0$ is the ground state energy. The inset shows the total spin during the optimization. (a) For the uniform Heisenberg chain, the ground state has $S^2 = 0$. We see that the equivariant gate can only find the ground state if we initialize the circuit with the Bell state $\ket{\Psi_-} = \frac{1}{\sqrt{2}}(\ket{01} - \ket{10})$, i.e., in the symmetry sector $S=0$. The horizontal-gate ansatz easily converges to a solution for both $S=0,1$ Bell states and the computational basis state $\ket{0}$, but struggles somewhat when initialized with a Haar random state. (b) For a Heisenberg chain with random couplings between the sites, the equivariant gate cannot find the ground state, which has non-integer spin. The horizontal gate, on the other hand, can find a high-quality approximation.}
    \label{fig:compare}
\end{figure*}
\section{Starting from a symmetry: Homogeneous spaces\label{sec:horizontal}}
The most straightforward way to construct a horizontal quantum gate is to start with symmetries given by a closed subgroup $K$ of $G$. The group $K$ comes with a Lie algebra $\kk$, which allows us to find the subspace $\mm$ and construct the gate via \Cref{eq:U_mm}. Since we are working with quantum gates, we usually set $G = \SU(N)$, although there is no reason why we could not pick other compact Lie groups such as $\SO(N)$ or $\Sp(N)$ and their products for $G$.

\subsection{The Bloch Sphere: Part I \label{ex:su2_u1_I}}
    The simplest non-trivial example of a horizontal gate is the case $G=\SU(2)$ and $K=\U(1)$ acting on a single-qubit subsystem, i.e., the symmetry is a global phase. If we take $\g = i\:\Span\{X,Y,Z\}$ as a representation of $\su(2)$, we can choose 
    \begin{align}
        \kk=i\:\Span\{Z\}\label{eq:bloch_kk}
    \end{align}
    as a representation of $\mathfrak{u}(1)$, which gives 
    \begin{align}
        \mm = i\:\Span\{X, Y\}\label{eq:bloch_mm}
    \end{align}
    as orthogonal complement. Note that $\kk_o=0$, since the commutant $\g^\kk = i\:\Span\{Z\}$, hence the equivariant tangent equals the tangent of the symmetry $K$. In the parameterization of \Cref{eq:U_mm}, we get 
    \begin{align}
        U(\btheta) = \exp{i(\theta_1 X + \theta_2 Y)},\label{eq:xy_gate}
    \end{align}
    and from \Cref{eq:commute_through}, we have
    \begin{align*}
        U'(\btheta) = \exp{ie^{-i\phi Z}(\theta_1 X + \theta_2 Y) e^{i\phi Z}}.
    \end{align*}
    This can be reparameterized as 
    \begin{align*}
        U'(\btheta) \equiv U(\btheta')=\exp{i(\theta_1' X + \theta_2' Y)},
    \end{align*}
    with
    \begin{align}
        \begin{pmatrix}
            \cos(2\phi) & -\sin(2\phi)\\
            \sin(2\phi) & \cos(2\phi)
        \end{pmatrix}
        \begin{pmatrix}
            \theta_1\\
            \theta_2
        \end{pmatrix} =
        \begin{pmatrix}
            \theta_1'\\
            \theta_2'
        \end{pmatrix}.\label{eq:reparam}
    \end{align}
    Hence the symmetry can be pulled through the gate at the cost of a reparameterization within $\mm$. Note that the action of the symmetry never results in the generation of the Pauli $Z$ in the exponent, which is a manifestation of \Cref{eq:comm_Amm}.

\subsection{\texorpdfstring{$SU(2)$}{SU(2)}-symmetric Hamiltonians}\label{ex:su4_su2_1_2}
    A more non-trivial example is the case $G=\SU(4)$ and $K=\SU(2)$, which has been considered in Refs.~\cite{nguyen2022theory, wierichs2023symm, east2023all}. This corresponds to an $\SU(2)$ symmetry subgroup on a two-qubit system. We use the tensor representation of $\su(2)$ to obtain 
    \begin{align*}
        \kk = i\:\Span\{XI +IX, YI + IY, ZI + IZ\},\nonumber
    \end{align*}
    which has $\zz(\kk)=0$. From the structure of $\kk$ we see that the commutant must be given by the one-dimensional subspace $i\:\Span \{\mathrm{SWAP}\}$---a fact which can also be deduced from the Schur--Weyl duality \cite{nguyen2022theory, ragone2022representation, east2023all}---hence there is only  a single equivariant direction. It is straightforward to find a basis for $\mm$ by choosing a basis of Pauli strings for $\g$ and calculating $\kk^\perp$ within $\g$ (see App.~\ref{app:decomp}). This gives
    \begin{align*}
     \mm=i\,\Span\{&XX,XY,XZ,YX,YY,YZ, ZX,ZY,ZZ, \\
     &XI-IX, YI-IY, ZI - IZ\},
    \end{align*}
    which contains $12$ elements in accordance with the dimension of $G/K$. In contrast to the single generator for equivariant gates, an arbitrary horizontal gate is thus parameterized by 12 parameters.
    
    The global $\SU(2)$ symmetry is of particular interest in the context of ground state searches of quantum many-body Hamiltonians such as the spin-$1/2$ Heisenberg chain,
    \begin{align}
        H =\frac{1}{4}\sum_{ i=1}^{n_q} \bm{S}_i\cdot \bm{S}_{i+1},\label{eq:heisenberg_uniform}
    \end{align}
    where $\bm{S} = (\sigma^x, \sigma^y, \sigma^z)$. This Hamiltonian has a global $\SU(2)$ symmetry~\cite{moudgalya2022hilbert}, which means that it commutes with the spin operators $S_{\alpha} = \sum_i^{n_q} \sigma_i^\alpha$, for $\alpha=x,y,z$, and has conserved charges $S_\alpha^2$.
    Specific $\SU(d)$-symmetric circuits ($d\geq2)$ were proposed in~\cite{seki2020symmadapted}, and a detailed construction based on the representation theory of $\SU(d)$ was proposed in~\cite{zheng2023geqml}. These cases were generalized and connected to spin networks~\cite{east2023all} and further described in the context of equivariant QML in~\cite{nguyen2022theory, meyer2023gqml}.

    An important factor in all these works is the initialization of the circuit. We know that for bipartite lattices the ground state of the anti-ferromagnetic Heisenberg model is always in the $S=0$ sector~\cite{lieb1962totalspin}. Initializing the circuit in the correct $S$-symmetry sector is crucial for geometric circuits to work, since they parametrize rotations within the specific spin subspace. However, there are Hamiltonians for which the spin subspace of the ground state is not known. For example, if we consider a Hamiltonian such as the random Heisenberg chain
    \begin{align}
        H =\frac{1}{4}\sum_{ i=1}^{n_q} h_{i}\:\bm{S}_i\cdot \bm{S}_{i+1},\label{eq:heisenberg_random}
    \end{align}
    with $h_{i}\sim \mathcal{N}(0,1)$, then the spin sector of the ground state is unknown. This lack of knowledge of the ground-state spin sector significantly inhibits the power of an equivariant quantum circuit, which relies on an initial state in the correct spin sector.
    
    In~\Cref{fig:compare}, we contrast circuits consisting of equivariant gates with circuits made from horizontal gates, both based on the above $\SU(2)$ symmetry. Specifically, we solve the VQE optimization problem~\cite{peruzzo2014vqe}:
    \begin{align}
        \min_\btheta E(\btheta) =  \bra{\psi_0}U^\dag(\btheta_{1:L}) H U(\btheta_{1:L})\ket{\psi_0} ,\label{eq:vqe}
    \end{align}
    where $U^\dag(\btheta_{1:L})$ consists of a bricklayer structure of gates (see \Cref{fig:brick}) and $\ket{\psi_0}$ is the initial state of the circuit.
    
    We find that horizontal quantum gates can circumvent the restrictions of equivariant gates and reliably find the ground state of both \Cref{eq:heisenberg_uniform} and \Cref{eq:heisenberg_random}, without prior knowledge of the ground-state spin sector. In \Cref{app:more_hom}, we generalize this construction to spin-3/2 symmetries, for which no equivariant circuits exist.

\section{Starting from an involution: symmetric spaces\label{sec:symmetric}}
Up to this point, we have not been concerned with how to apply the gate in \Cref{eq:U_mm} in terms of standard quantum gates; we have simply assumed some gate decomposition scheme that implements the gate for us. We know that such a decomposition can always be constructed with general unitary decomposition schemes~\cite{khaneja2001cartan}, but these can be very inefficient. Here, we focus on a particular type of reductive homogeneous space called a \emph{symmetric space}. Such spaces correspond to a homogeneous space $G/K$ with additional structure that allows one to decompose a horizontal gate into smaller constituents that are easier to implement.

Instead of starting from a symmetry $K$, we start with an involution $\varphi\colon\g \to\g$, which is an automorphism of $\g$ with the property $\varphi^2 = I$. Then $\varphi$ splits $\g$ into a $+1$ ($\kk$) and $-1$ ($\mm$) eigenspace as in~\Cref{eq:decomp}. Such a split is called a \emph{Cartan decomposition} and comes with a (non-unique) maximal Abelian subalgebra $\hh\subset \mm$, which is called a Cartan subalgebra. The additional structure of a symmetric space makes it possible to classify them (see~\Cref{sec:classification} and \cite{helgason1979differential}).

In addition to the commutation relations in~\Cref{eq:homogeneous}, symmetric spaces have the property that
\begin{align*}
    [\mm,\mm]\subseteq \kk. 
\end{align*}
An important consequence of this property is that we have
\begin{align}
    e^{\mm} = \{ k e^h k^{-1} \,|\, k\in K, \, h\in\hh \}
    \label{eq:kak},
\end{align}
which is known as the \emph{KAK theorem} (see, e.g., \cite[Chapter VII]{knapp2013lie}). It implies that any element of $\mm$ can be reached via the adjoint action of $K$ on the Cartan subalgebra $\hh$. This theorem has been been used extensively in the literature~\cite{vatan2004opt2qubit,khaneja2001cartan, dalessandro2006decompuni, bullock2004note, bullock2004canonical, dallesandro2007quantum, kokcu2022fixed} to decompose arbitrary unitaries into products of one- and two-qubit quantum gates. Here, we can make use of it to explicitly decompose horizontal gates into simpler one- and two-qubit gates.

\subsection{The Bloch Sphere: Part II\label{ex:su2_u1_II}}

The Bloch sphere example in \Cref{ex:su2_u1_I} is a symmetric space, which can be shown by considering the involution,
\begin{align*}
    \varphi(A) = -X A^T X.
\end{align*}
We see that under this involution, $iZ$ stays positive $\varphi(iZ)=iZ$, whereas $\theta(iX)=-iX$ and $\theta(iY)=-iY$. Hence we can identify the positive and negative eigenspaces of $\varphi$ as $\kk$ and $\mm$ in Eqs.\ \eqref{eq:bloch_kk} and \eqref{eq:bloch_mm}, respectively.

\begin{figure}[htb!]
    \centering
    \begin{quantikz}
        & \gate{R_z(\theta_1)}\gategroup[1,steps=1,style={dashed,rounded
        corners,fill=MyRed!60, inner
        xsep=1pt},background,label style={label
        position=below,anchor=north,yshift=-0.2cm}]{$\kk$}  &  \gate{R_y(\theta_2)}\gategroup[1,steps=1,style={dashed,rounded
        corners,fill=MyBlue!60, inner
        xsep=1pt},background,label style={label
        position=below,anchor=north,yshift=-0.2cm}]{$\hh$}  & \gate{R_z^\dag(\theta_1)}\gategroup[1,steps=1,style={dashed,rounded
        corners,fill=MyRed!60, inner
        xsep=1pt},background,label style={label
        position=below,anchor=north,yshift=-0.2cm}]{$-\kk$} 
    \end{quantikz}
    \caption{\textbf{Circuit decomposition of a one-qubit gate using the KAK theorem.} The gate of \Cref{eq:xy_gate} can be decomposed into three separate one-qubit quantum gates. The number of free parameters corresponds to the dimension of $G/K$, which is $\dim(\SU(2))- \dim(\U(1)) = 2$.}
    \label{fig:su2_gate}
\end{figure}
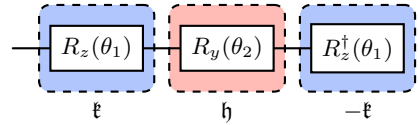

We choose $\hh= i\:\Span\{Y\}$, which is clearly a maximal Abelian subalgebra of $\mm$, and using \Cref{eq:kak}, we obtain
\begin{align*}
    U(\btheta) = e^{i\theta_1 Z} e^{i\theta_2 Y} e^{-i\theta_1 Z}
\end{align*}
as a decomposition of the gate in \Cref{eq:xy_gate}; this decomposition is straightforward to implement in practice (see \Cref{fig:su2_gate}).

\subsection{Single-Qubit Orbits\label{ex:su4_su2_su2}}
Consider the case of $\g=\su(4)$, with the involution
\begin{align*}
    \varphi(A) = -(YY) A^T (YY).
\end{align*}
This splits the Lie algebra as $\g = \kk\oplus\mm$ with
\begin{align*}
    \kk &= i\:\Span\{XI, YI, ZI, IX, IY, IZ\},\\
    \mm &= i\:\Span\{XX, YY, ZZ, XY, XZ, YX, YZ, ZX, ZY\},\\
    \hh &= i\:\Span\{XX, YY, ZZ\},
\end{align*}
as bases for $\kk$, $\mm$ and $\hh$, which is the choice of Refs.~\cite{khaneja2001cartan,bullock2004note}. We see that 
\begin{align*}
\kk = \su(2)\otimes I + I \otimes \su(2) \cong \su(2)\oplus\su(2) \cong \so(4),
\end{align*}
and we have the symmetric space
\begin{align*}
    G/K = \SU(4)/\SO(4).
\end{align*}
Since the product group $\SU(2)\times\SU(2)$ is a double cover of $\SO(4)$, we can also consider the homogeneous space $\SU(4)/(\SU(2)\times\SU(2))$.

\begin{figure}[htb!]
    \resizebox{\columnwidth}{!}{
    \begin{quantikz}
        & \gate{R_3(\bm\alpha_1)}\gategroup[2,steps=1,style={dashed,rounded
        corners,fill=MyRed!60, inner
        xsep=1pt},background,label style={label
        position=below,anchor=north,yshift=-0.2cm}]{$\kk$}  &
        \targ{1}\gategroup[2,steps=5,style={dashed,rounded
        corners,fill=MyBlue!60, inner
        xsep=1pt},background,label style={label
        position=below,anchor=north,yshift=-0.2cm}]{$\hh$} & \gate{R_z(\theta_1)}& \ctrl{1} & \qw &\targ{1} & \gate{R_3^\dag(\bm\alpha_1)}\gategroup[2,steps=1,style={dashed,rounded
        corners,fill=MyRed!60, inner
        xsep=1pt},background,label style={label
        position=below,anchor=north,yshift=-0.2cm}]{$-\kk$}  & \\
        & \gate{R_3(\bm\alpha_2)} &\ctrl{-1} &\gate{R_y(\theta_2)} & \targ{-1} & \gate{R_y(\theta_3)} & \ctrl{-1} & \gate{R_3^\dag(\bm\alpha_2)} &
    \end{quantikz}
    }
    \caption{\textbf{Circuit decomposition of a two-qubit horizontal gate using the KAK theorem.} The homogeneous space $\SU(4)/(\SU(2)\times \SU(2))$ has dimension $\dim(\SU(4)) - 2\dim(\SU(2)) = 9$, which is reflected by the number of parameters used in the circuit decomposition. The gate $R_3$ corresponds to an arbitrary $\SU(2)$ rotation.}
    \label{fig:vatan}
\end{figure}
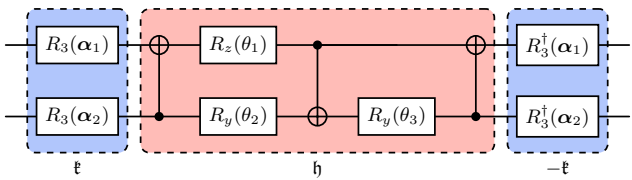

The KAK theorem~\Cref{eq:kak} then tells us that we can implement the horizontal gate $U_\mm(\btheta)$ as
\begin{align*}
    U_\mm(\bm{\alpha},\bphi) = e^{iB(\bm{\alpha})} e^{ih(\bphi)} e^{-iB(\bm{\alpha})},
\end{align*}
where
\begin{align*}
    B(\bm{\alpha}) &= \alpha_1 XI+ \alpha_2 YI+ \alpha_3 ZI +\alpha_4  IX +\alpha_5  IY+\alpha_6  IZ,\\
    h(\bphi) &= \theta_1 XX + \theta_2  YY + \theta_3 ZZ.\nonumber
\end{align*}
These exponentials can be implemented as a product of simpler gates with the method of Vatan~\cite{vatan2004opt2qubit}, which describes the optimal circuit decomposition of an $\SU(4)$ gate in terms of CNOTs. The one difference is that the parameters in the first and last layers are equal to each other, whereas for a general $\SU(4)$ gate, they are different (see~\Cref{fig:vatan}). 

\section{Starting from quantum states: stabilizing gates}\label{sec:stabilizing}

We have seen how to construct horizontal quantum gates from symmetries and involutions. Here, we consider a construction that starts with a quantum state or a set of quantum states. We start by investigating the action of a Lie group $G$ on the Hilbert space $\mathcal{H}$.
Given a subspace of states $S\subset\mathcal{H}$, we define a closed subgroup of $G$ called the \emph{stabilizer} of $S$,
\begin{align*}
    K=G_S =\{g\in G\,|\, g \cdot S = S\},
\end{align*}
which contains all elements $g\in G$ that send the states from $S$ into $S$. We can use the stabilizer to construct the homogeneous space $G/K$, identify the Lie algebra $\kk$ that generates $K$ and construct $\mm$ as the orthogonal complement within $\g$. We will consider three examples that motivate a construction of this type.

\subsection{The Bloch Sphere: Part III\label{ex:su2_u1_III}}

\begin{figure}[htb!]
    \centering
    \includegraphics[width=\columnwidth]{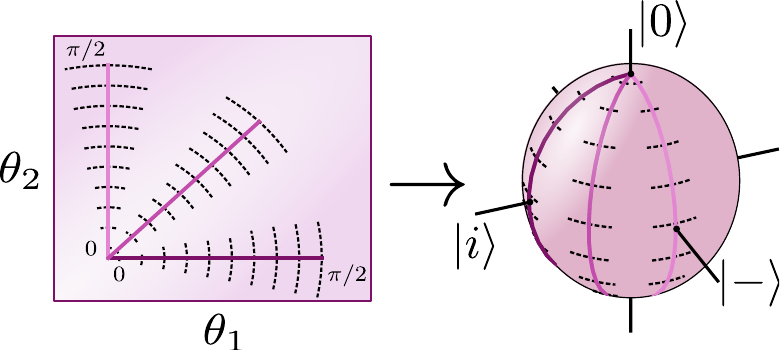}
    \caption{\textbf{Visualising the action of a horizontal gate on the Bloch sphere}. The orbits under the $\U(1)$ symmetry correspond to circles around the $\ket{0}$ axis of the Bloch sphere. We see that the horizontal gate always moves the states in the direction $\mm$ (colored lines) orthogonal to the symmetry direction (dashed lines). The three colored lines show the paths $U(t) = e^{itX}$, $U(t) = e^{itY}$ and $U(t) = e^{\frac{it}{\sqrt{2}}(X+Y)}$ for $t\in[0,\pi/2]$. The symmetry directions are generated by the action of the stabilizer in \Cref{eq:stab_bloch} on the states of the three colored lines for the values $\phi\in [-0.1,0.1]$. }
    \label{fig:bloch_cosets}
\end{figure}

We consider the stabilizer of the single-qubit coset $[\ket{0}]$, which is determined by the equivalence relation $\ket{0}\sim e^{i\phi}\ket{0}$. We find the stabilizer to be
\begin{align}
    G_{[\ket{0}]} = \left\{\begin{pmatrix}
        e^{i\phi} & 0\\
        0 & e^{-i\phi}
    \end{pmatrix} \,\Big|\, \phi\in\R \right\}
    = \left\{ e^{i\phi Z} \,\big|\, \phi\in\R \right\},\label{eq:stab_bloch}
\end{align}
which is a subgroup of $G=\SU(2)$.
Hence $K=\U(1)$ and $\kk = i\:\Span\{Z\}$, just as in \Cref{eq:bloch_kk}. We can investigate how the corresponding horizontal gate $e^{i(\theta_1 X + \theta_2 Y)}$ acts on the coset. 
Using the reparameterization of \Cref{eq:reparam}, we find
\begin{align*}
    e^{i(\theta_1 X + \theta_2 Y)} & e^{i\phi Z}\ket{0} \\ 
    &= e^{i\phi Z}
        \left(\cos (r)\ket{0}+
        (i \theta_1'-\theta_2') \frac{\sin(r)}{r}\ket{1}\right)
\end{align*}
with $r = \sqrt{(\theta_1')^{2}+(\theta_2')^2}$. In \Cref{fig:bloch_cosets}, we show the resulting states on the Bloch sphere for different parameters $\theta_1, \theta_2$ and $\phi$. We see that each state can be reached from $[\ket{0}]$. 

\subsection{Efficient Direct State Transformations\label{ex:su_4_u3}}

Here we generalize the idea in \Cref{ex:su2_u1_III}, to a 2-qubit system. We again fix a coset $[\ket{0}]$ and find its stabilizer inside $G=\SU(4)$, given by
\begin{align*}
    G_{[\ket{0}]} = \left\{\begin{pmatrix}
        e^{i\phi} & 0\\
        0 & e^{A}
        \end{pmatrix} \,\Big|\, \phi\in\R,\, A\in\uu(3),\, \Tr{A}=-i\phi \right\}. 
\end{align*}
We see that $K=G_{[\ket{0}]}=\mathrm{S}(\U(1)\times\U(3))\cong \U(3)$. It is quickly verified that the corresponding Lie algebra and horizontal subspace are: 
\begin{align*}
    \kk&=\left\{\begin{pmatrix}
        i\phi & 0\\
        0 & A
    \end{pmatrix} \,\Big|\, \phi\in\R,\, A\in\uu(3),\, \Tr{A}=-i\phi \right\},\\
    \mm &= \left\{\begin{pmatrix}
    0 & \bm{x}\\
    -\bm{x}^\dag& 0
    \end{pmatrix} \,\Big|\, \bm{x}\in\mathbb{C}^3 \right\}.
\end{align*}
Note that the resulting $\SU(4)/\U(3)$ horizontal gate has $6$ real parameters, which equals the number of real parameters we need to describe a 2-qubit state (up to a global phase and normalization). On the other hand, a quantum gate that fully parameterizes $\SU(4)$ would have $\dim(\SU(4)) = 15$ parameters. More generally, we have $\dim(\SU(N)/\U(N-1))=2N-2$ as opposed to the $N^2-1$ of a $\SU(N)$ gate, which provides a quadratic reduction in the number of parameters. We note that while the above gate is defined with respect to the $\ket{0}$ state, it can act on any input state. 

\begin{figure}[htb!]
    \centering
    \subfloat[GUE]{\includegraphics[width=0.45\textwidth]{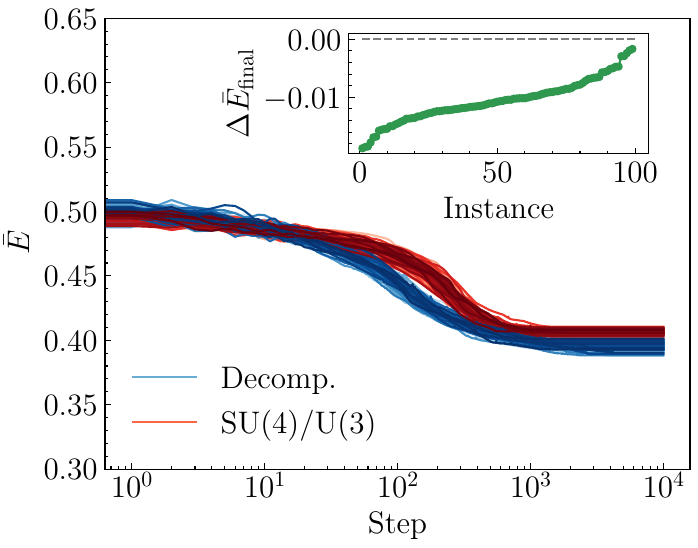}}
    \hspace{5mm}
    \subfloat[GOE]{\includegraphics[width=0.45\textwidth]{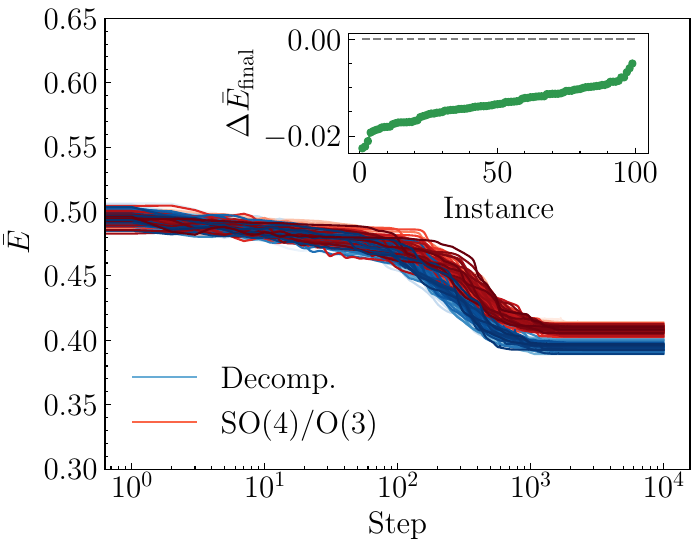}}
    \caption{\textbf{Optimization trajectories of VQE with bricklayer circuits on systems of size $n_q=12$ and circuits with depth $L=12$.} We compare the performance of bricklayer circuits (see \Cref{fig:brick}) consisting of general \textbf{(a)} $\SU(4)$ gates with horizontal $\SU(4)/\U(3)$ gates and \textbf{(b)} $\SO(4)$ gates with horizontal $\SO(4)/\mathrm{O}(3)$ gates. The decomposed gates correspond to the optimal decompositions of~\cite{vatan2004opt2qubit}. We show the difference of the relative errors in energy $\Bar{E} = (E - E_{\min}) / (E_{\max} - E_{\min})$ for both the decomposed gates and the horizontal gates, where $E_{\max}$ and $E_{\min}$ are the largest and smallest eigenvalues of the target Hamiltonian, respectively. In the inset, we show $\Delta \bar{E}_{\mathrm{final}} = \Bar{E}_{\mathrm{Hor}.} - \Bar{E}_{\mathrm{Decomp.}}$.
    For (a), the target Hamiltonian is a randomly generated Hermitian operator sampled from the Gaussian unitary ensemble (GUE), whereas in (b) the target is a random orthogonal matrix sampled from the Gaussian orthogonal ensemble (GOE). For both cases, we see that the horizontal gates can achieve similar performance (inset) with $6$ parameters per gate instead of $15$. This results in a shorter overall optimization time, since we have less parameters to optimize per step. From the insets, we see that horizontal gates perform slightly worse, since $\Delta \bar{E}_{\mathrm{final}}$ is negative.
    }
    \label{fig:trajectories}
\end{figure}

One caveat of this horizontal gate is that there is a clash between the structure of the Lie algebra as a direct sum ($\kk=\uu(1)\oplus\su(3)\cong\uu(3)$), versus the tensor product structure of the Hilbert space that the gate is acting on ($\mathcal{H}=\mathbb{C}^2\otimes\mathbb{C}^2$). A basis for $\kk$ in terms of Pauli strings would look quite involved, and even though $\SU(4)/\U(3)$ is a symmetric space it is not straightforward to construct a natural circuit decomposition via the KAK theorem.
Despite the fact that there is no clear simplified description of this gate in terms of standard gates, we can investigate its usefulness in practice. In~\Cref{fig:trajectories}, we investigate how the gate $\SU(4)/\U(3)$ and its real counterpart $\SO(4)/\mathrm{O}(3)$ perform in a standard bricklayer circuit, when compared to a general $\SU(4)$ and $\SO(4)$ gates, respectively. We generate random skew-symmetric and symmetric Hamiltonians $H$ and solve the VQE problem of \Cref{eq:vqe}. We find that we can achieve similar performance with horizontal gates compared to their full $\SU$ or $\SO$ counterparts, with quadratically fewer parameters. 
\subsection{Further Generalizations}
The above technique can be used to more efficiently parameterize unitaries when the input state is fixed—a commonly encountered situation—and works in arbitrary dimension. In App.\ \ref{app:stab} we outline the general construction for horizontal gates of the form $\SU(N)/\U(N-1)$; we also discuss there a generalization to Grassmannians and gates acting on a charge-preserving subspace.

\section{Conclusion}

We have presented a method to incorporate continuous symmetries into quantum circuits by constructing quantum gates that only act on horizontal subspaces. We have shown that these gates are powerful and can be used to construct gates for a variety of different continuous symmetries. Moreover, the proposed gates can handle cases for which an equivariant circuit may be too restrictive or does not exist. Additionally, we have shown that the $\SU(4)/\U(3)$ gates we considered are significantly more efficient when compared to general $\SU(4)$ gates. 
Based on these observations,
we believe that any variational algorithm that uses general $\SU(N)$ gates would benefit from the $\SU(N)/\U(N-1)$ gates, because the number of parameters will be reduced. Generalizing this
perspective, we suggest that horizontal quantum
gates can be used to reduce circuit complexity
and the number of parameters whenever a symmetry
is respected.
Building upon the theoretical foundations we have established here, further studies can be conducted to test the efficacy of horizontal gates for specific systems of interest.

For the numerical results in this work, we have mostly relied on the exponential description in \Cref{eq:gate}. Decomposing these gates into smaller circuits is made easier when $G/K$ is a symmetric space, since we can then rely on the KAK theorem. However, the resulting blocks of gates can still be expensive to implement with a standard gate set (see the additional examples in \Cref{sec:symmetric}). 

\clearpage

\noindent A possible future direction would be to explore alternative decomposition strategies for the gates we considered, so that horizontal gates could be compiled more easily to the native gate set of current quantum hardware. 

The list of homogeneous spaces that we explored is by no means exhaustive and future studies will explore how more exotic symmetries could be implemented using horizontal gates. For example, recent work on non-Abelian gauge theories could benefit from symmetry-respecting gates~\cite{klco2020nonabelian, atas2021su2}. Additionally, although we excluded the exceptional Lie groups from our presentation, there are homogeneous spaces of exceptional groups that show up in Grand Unification Theories~\cite{gross1989heterotic}. A further abstract extension of our work would be the investigation of double coset spaces $H\backslash G /K$, where a coset is composed of an element $HgK$, i.e., it is multiplied both on the left and right by different groups. The applications of such gates are not clear right now, since double coset spaces are primarily of mathematical interest.

Finally, the Riemannian geometry of homogeneous spaces has been extensively studied, so it would be interesting to connect the geometry of these spaces, and their optimization properties, to variational quantum circuits. 
Furthermore, we note that the one could define Riemannian gradient flows over the homogeneous-space quantum gates~\cite{wiersema2023riemann}; there exists prior work in this direction in the quantum control setting~\cite{SchulteHerbruggen2010gradflow, Helmke1994optdyn}, and these ideas could be applied in variational quantum computing to directly optimize circuits. 

\section{Acknowledgements}

We thank David Wierichs, Korbinian Kottmann, Martin Larocca, and Efekan K\"okc\"u for valuable discussions.
AFK acknowledges financial support from the National Science Foundation under award No. 1818914: PFCQC: STAQ: Software-Tailored Architecture for Quantum co-design and No. 2325080:
PIF: Software-Tailored Architecture for Quantum Co-Design (STAQ II).
BNB was supported in part by a Simons Foundation grant No. 584741.

Resources used in preparing this research were provided, in part, by the Province of Ontario, the Government of Canada through CIFAR, and companies sponsoring the Vector Institute \url{https://vectorinstitute.ai/#partners}.
\bibliography{library}
\clearpage
\renewcommand{\appendixtocname}{Appendices}
\onecolumngrid

\renewcommand\thefigure{\Alph{section}\arabic{figure}}  
\renewcommand\thetable{\Alph{section}\Roman{table}}  
\setcounter{figure}{0}
\setcounter{table}{0}
\renewcommand\theHfigure{S\arabic{figure}}
\renewcommand\theHtable{S.\Roman{table}}

\renewcommand{\thesection}{\Alph{section}}
\renewcommand{\thesubsection}{\Roman{subsection}}

\renewcommand{\theequation}{\Alph{section}\arabic{equation}}
\counterwithin*{equation}{section}
\setcounter{section}{0}
\pagebreak
\begin{center}
\textbf{\large Appendices}
\end{center}
\setcounter{equation}{0}
\setcounter{page}{1}
\makeatletter

\section{The tangent space of a homogeneous space \label{app:tang_hom}}
For a compact Lie group $G$ and closed subgroup $K$, there exists a projection $\pi\colon G\to G/K$ with $\pi(g) = gK$ for $g\in G$, so that $G/K$ is a homogeneous space. In this case, $\pi$ is a \emph{submersion}, a smooth map whose differential at every point is onto~\cite[Proposition 4.1]{arvanitogeorgos2003introduction}. Note that $\pi(k) = k K = K$ for all $k\in K$, and so the projection $\pi$ is constant when restricted to $K$. Hence, for $k\in K$ and $X\in\kk$, we have
\begin{align}
    d\pi_k(X)= \frac{d}{dt}\pi \bigl(k\exp{t X} \bigr)\Big|_{t=0} = \frac{d}{dt} K \Big|_{t=0} = 0.\label{eq:ker_k}
\end{align}
Therefore, $d\pi_k(\kk) = 0$. 
The differential at the identity element $e\in G$ is a surjective map
$d\pi_e\colon \g \to T_{K} (G/K)$. From \Cref{eq:ker_k}, we know that $\ker(d\pi_e)=\kk$; hence, we can take the quotient of the vector spaces $\g$ and $\kk$ to obtain the tangent space of $G/K$ at the point $\pi(e)=K$:
\begin{align*}
    T_{K}(G/K)\cong \g/\kk \cong \mm.
\end{align*}
In the case where $G$ is compact, we have that $\mm = \kk^\perp$ \cite{arvanitogeorgos2003introduction}.

\section{Lie algebra decomposition\label{app:decomp}}
In this section, we outline how decomposing the Lie algebra into the four subspaces of \Cref{eq:wierichs_decomp} can be achieved. We provide Python code at \cite{our_code} for these procedures. We restate the Lie algebra decomposition here:
\begin{align*}
    \g =\lefteqn{\underbrace{\phantom{\mathfrak{r} \oplus\g^\kk_o}}_{\mm}}\mathfrak{r} \oplus
    \lefteqn{\overbrace{\phantom{\g^\kk_o\oplus\zz(\kk)}}^{\g^\kk }}\g^\kk_o \oplus
    \lefteqn{\underbrace{\zz(\kk) \oplus \kk_o}_{\kk}.}
\end{align*}
We assume that we have orthonormal bases $\{X_i\}$ and $\{Y_j\}$ for $\g$ and $\kk$, respectively, with $\dim(\g)=N$ and $\dim(\kk)=M$. 
 
\subsection*{Finding  \texorpdfstring{$\mm$}{}}
We can find $\mm$ by finding the orthogonal complement of $\kk$ within $\g$. Remember that for the Lie algebras considered in our work, $\g$ and $\kk$ as vector spaces can be identified with $\R^N$ and $\R^M$, respectively. Then the orthonormal basis $\{X_i\}$ becomes the standard basis $\{e_i\}$, and the trace inner product between $A=\sum_i a_i X_i$ and $B=\sum_i b_i X_i$ reduces to the standard inner product:
\begin{align*}
    \Tr{A^\dag B} &= \sum_{i,j=1}^N\Tr{a_i X_i^\dag b_j X_j}\\
    &=\sum_{i=1}^N a_i b_i \equiv  \bm{a}\cdot \bm{b},
\end{align*}
with $\bm{a}=(a_1,\ldots, a_N)^T$, $\bm{b}=(b_1,\ldots, b_N)^T \in\R^N$. 
In this way, we identify the vector $A\in\g$ with its coordinates $\bm{a}\in\R^N$ relative to the basis $\{X_i\}$.
We can now construct the orthogonal complement to $\kk$ with a standard result from linear algebra. Let 
us express each of the basis vectors of $\kk$ in terms of the basis of $\g$:
\begin{align*}
Y_j = \sum_{k=1}^N a_{jk} X_k, \qquad j=1,\dots,M.
\end{align*}
Then the kernel of the $M\times N$ matrix $A = (a_{jk})$ is exactly $\kk^\perp=\mm$.

\subsection*{Finding  \texorpdfstring{$\g^\kk$}{}}
To find the commutant $\g^\kk$ of $\kk$ in $\g$, we can use the following approach. Recall that the commutant is given by
\begin{align*}
    \g^\kk = \{X \in \g \,|\, [X,Y]=0,\: \forall Y\in\kk\}. 
\end{align*}
The adjoint representation $\ad_X\colon\g\to\g$ of an element $X \in \g$ is given by the linear map
\begin{align*}
    \ad_{X} = [X, \:\cdot\:].
\end{align*}
For a fixed $X$, we can explicitly represent this map as an $N\times N$ matrix relative to the basis $\{X_i\}$. The condition $[X,Y]=0,\: \forall Y\in\kk$ thus comes down to finding the joint kernel of all $\ad_Y$ with $Y\in\kk$. By linearity, it is enough to do this for all $Y=Y_j$:
\begin{align*}
    \g^\kk = \bigcap_{j=1}^M \ker\left(\ad_{Y_j}\right),
\end{align*}
which is straightforward to implement numerically. 
\subsection*{Finding  \texorpdfstring{$\g^\kk_o$}{} and  \texorpdfstring{$\zz(\kk)$}{}}
Once $\mm$ and $\g^\kk$ are found, we simply take the intersections of $\mm$ and $\kk$ with $\g^\kk$ to get
\begin{align*}
    \g^\kk_o &= \g^\kk \cap \mm = \g^\kk \cap \kk^\perp,\\
    \zz(\kk) &= \g^\kk \cap \kk = \g^\kk \cap \mm^\perp.
\end{align*}
One way to find a basis for $\g^\kk_o$ is to view it as the set of all vectors from $\g^\kk$ that are orthogonal to all vectors from $\kk$, or equivalently to the basis vectors $Y_j$, similarly to how we found $\mm=\kk^\perp$ above.
Finally, if needed, we can determine $\mathfrak{r}$ as the orthogonal complement of $\g^\kk_o$ in $\mm$, and $\kk_o$ as the orthogonal complement of $\zz(\kk)$ in $\kk$.

\subsection*{Finding \texorpdfstring{$\hh$}{}}
For a symmetric space, we are interested in finding a maximal Abelian subalgebra $\hh\subset \mm$. Constructing a basis for $\hh$ can be achieved by iteratively adding elements so that they all commute. We achieve this as follows. We start by picking a nonzero element $h_1\in \mm$. We then calculate the kernel of $\ad_{h_1}$ on $\mm$, which contains all elements of $\mm$ that commute with $h_1$:
\begin{align*}
    \ker(\ad_{h_1}|_{\mm}) = \mm\cap \ker(\ad_{h_1}).
\end{align*}
From this vector space, we select an element $h_2$ that is not in the span of $\{h_1\}$. 
Note that, by construction, $[h_1,h_2]=0$.
We repeat the above procedure for the set $\{h_1, h_2\}$, calculating the common kernel of $\ker(\ad_{h_1})$ and $\ker(\ad_{h_2})$ on $\mm$:
\begin{align*}
    \ker(\ad_{h_1}|_{\mm}) \cap \ker(\ad_{h_2}|_{\mm}).
\end{align*}
From this vector space, we pick a new element $h_3 \not\in\Span\{h_1, h_2\}$. By construction, $h_1,h_2,h_3$ commute and are linearly independent. We keep repeating this until we we can no longer increase the span of $\{h_1,h_2,\dots\}$, which indicates that it is a maximal Abelian subalgebra in $\mm$. Moreover, the elements $\{h_1,h_2,\dots\}$ form a basis for $\hh=\Span\{h_1, h_2,\dots\}$.

\section{Classification of symmetric spaces}
\label{sec:classification}
The classification of symmetric spaces can be found in great detail in \cite[Chapter X.3]{helgason1979differential}. Here we simply tabulate the result of this classification for the cases where the symmetric space is a quotient of compact matrix Lie groups. To describe the involutions that split the Lie algebra into $\g = \kk \oplus \mm$, we need the following matrices:
\begin{align}
    I_{p,q} = \begin{pmatrix}
            -I_p & 0 \\
            0 & I_q\\
        \end{pmatrix},\quad 
    J_n = \begin{pmatrix}
            0 & I_n\\
            -I_n &0\\
        \end{pmatrix},\quad
    K_{p,q} = \begin{pmatrix}
            -I_p & 0& 0& 0 \\
            0 & I_q & 0 & 0\\
            0 & 0 & -I_p &0\\
            0 & 0 & 0 & I_q\\
        \end{pmatrix}\label{eq:matrices},
\end{align}
in addition to the $n\times n$ identity matrix $I_n$. In \Cref{tab:symmetric_spaces}, we tabulate the classical symmetric spaces.
\begin{table}[htb!]
    \centering
    \begin{tabular}{|c|c|c|c|c|c|}
        \hline
        Type & $\g$ & $\kk$ & $\mm$ & Involution & $G/K$  \\ \hline
        {AI}   & $\su(N)$  & $\so(N)$ & $i \:\mathrm{Sym}(N)$ &  $\varphi(A) = A^*$ &$\SU(N)/\SO(N)$ \\
        {AII}  & $\su(2N)$ & $\sp(N)$ & $\su(N)\oplus\so(N,\mathbb{C})$ & $\varphi(A) = J_n A^* J_n^{T}$&$\SU(2N)/\Sp(N)$\\
        {AIII} & $\su(p+q)$& $\mathfrak{s}(\mathfrak{u}(p)\oplus\mathfrak{u}(q))$&$\mathfrak{s}(\mathfrak{u}(p)\oplus\mathfrak{u}(q))^\perp $ & $\varphi(A) = I_{p,q} A I_{p,q}$ & $\SU(p+q)/\mathrm{S}(\mathrm{U}(p)\times \mathrm{U}(q))$\\
        {BDI}  & $\so(p+q)$& $\so(p)\oplus\so(q)$&$\so(p)^\perp \cap\so(q)^\perp $  &$
            \varphi(A) = I_{p,q} A I_{p,q}$ & $\SO(p+q)/\mathrm{SO}(p)\times \mathrm{SO}(q)$\\\
        {CI}   & $\sp(N)$  & $\uu(N)$ &$\mathfrak{u}(N)^\perp$ & $\varphi(A) = A^*$ & $\Sp(N)/\mathrm{U}(N)$\\
        {CII}  & $\sp(p+q)$ & $\sp(p)\oplus \sp(q)$ & $\sp(p)^\perp\cap \sp(q)^\perp$ & $
            \varphi(A) = K_{p,q} A K_{p,q} $ & $\Sp(p+q)/\Sp(p)\times \Sp(q)$\\
        {DIII} & $\so(2N)$& $\uu(N)$ &$\mathfrak{u}(N)^\perp$ & $\varphi(A) = J_n A J_n^{T} $ & 
        $\SO(2N)/\mathrm{U}(N)$\\   \hline     
    \end{tabular}
    \caption{\textbf{The classical symmetric spaces}. The labels of the types A, B, C, D 
    correspond to the classical Lie groups $\SU(N)$, $\SO(N)$ ($N$ odd), $\Sp(N)$ and $\SO(N)$ ($N$ even), respectively. For a matrix $A$, we denote by $A^*$ the matrix whose entries are the complex conjugates of the entries of $A$. In the first row, $\mathrm{Sym}(N)$ is the space of $N\times N$ real symmetric matrices.}
    \label{tab:symmetric_spaces}
\end{table}
For the interested reader, we also include a list of the exceptional symmetric spaces below:
\begin{align*}
    &\frac{E_6}{\Sp(4)},\quad \frac{E_6}{\SU(6)\times \SU(2)},\quad \frac{E_6}{F_4},\quad \frac{E_6}{\SO(10)\times \SO(2)}, \quad\frac{E_7}{\SU(8)}, \quad\frac{E_7}{\SO(12)\times \SU(2)}
    \\
    &\frac{E_7}{E_6\times \SO(2)}, \quad\frac{E_8}{\SO(16)},\quad \frac{E_8}{E_7\times \SU(2)}, \quad\frac{F_4}{\SO(9)}, \quad\frac{F_4}{\Sp(3)\times \SU(2)},\quad \frac{G_2}{\SO(4)}.
\end{align*}

\section{More examples \label{app:more}}
In this section, we include a list of increasingly esoteric examples of horizontal quantum gates. The Lie algebra decompositions reported in this section are determined with the methods outlined in \Cref{app:decomp}.
\subsection{Homogeneous Spaces\label{app:more_hom}}
\subsubsection{\texorpdfstring{$\SU(4)/\SU(2):$ A Spin-$3/2$ Example}{}}\label{ex:su4_su2_3_2}
    We can also choose different representations of $\SU(2)$ as a quotient.
    For example, we can consider the example $G=\SU(4)$ and $K=\SU(2)$ where we take the spin-$3/2$ representation of $\su(2)$, which is given by $\kk = i\:\Span\{S_x, S_y, S_z\}$ with
    \begin{align*}
        S_x &= \begin{bmatrix}
    0 & \sqrt{3} & 0 & 0 \\
    \sqrt{3} & 0 & 2 & 0 \\
    0 & 2 & 0 & \sqrt{3} \\
    0 & 0 & \sqrt{3} & 0 \\
    \end{bmatrix},\quad S_y = \begin{bmatrix}
    0 & -i\sqrt{3} & 0 & 0 \\
    i\sqrt{3} & 0 & -2i & 0 \\
    0 & 2i & 0 & -i\sqrt{3} \\
    0 & 0 & i\sqrt{3} & 0 \\
    \end{bmatrix},\quad S_z = \begin{bmatrix}
    3 & 0 & 0 & 0 \\
    0 & 1 & 0 & 0 \\
    0 & 0 & -1 & 0 \\
    0 & 0 & 0 & -3 \\
    \end{bmatrix}.
    \end{align*}
We have a trivial center $\zz(\kk)=0$ and the commutant is trivial as well $\g^\kk = 0$. Hence there is no possible equivariant quantum gate for this symmetry. However, we are still left with 12 horizontal directions in $\mm$ and so the horizontal quantum gate could be useful. We find
\begin{align*}
     \mm = i\,\Span\{&
     aIX - bXX  + cYY,
     aIY + bXY + cYX,
    -aIY + cXY + bYX , 
    -aIX -cXX + bYY ,
    -dIZ + aZI,\\
    & XI,XZ,YI , YZ ,ZX ,ZY ,ZZ 
    \},
\end{align*}
where the coefficients are $a= \frac{1}{\sqrt{5}}$, $b=\frac{5+\sqrt{15}}{10}$, $c = \frac{5-\sqrt{15}}{10}$ and $d=\frac{2}{\sqrt{5}}$.
\subsection{Symmetric Spaces}
\subsubsection{\texorpdfstring{$\SU(4)/\Sp(2):$ Exotic Symmetry I }{}}\label{ex:su4_sp2} 
We consider the example $G=\SU(4)$ and $K=\Sp(2)$, where $\Sp$ is the symplectic group. A basis for $\sp(2)$ is given by 
\begin{align*}
    \kk=i\:\Span\{IY,XI, XX, XZ, YI, YX, YZ, ZI, ZX, ZZ\},    
\end{align*}
which all satisfy the symplectic property $J_2A^TJ_2=A$ for $A\in\kk$
and $J_2=iYI$ as defined in \Cref{eq:matrices}. For the horizontal subspace we find
\begin{align*}
    \mm &= i\:\Span\{IX,IZ,XY,YY,ZY \}.
\end{align*}
Again, there is no equivariant subspace, since the center and commutant of $\kk$ are trivial. We note that $G/K$ is a symmetric space of type AII (see \Cref{tab:symmetric_spaces}). A maximal Abelian subalgebra is $\hh = i\:\Span\{IX\}$.

\subsubsection{\texorpdfstring{$\SO(4)/\mathrm{U}(2):$ Exotic Symmetry II}{}}\label{ex:so4_u2}

A more contrived example is the space $\SO(4) / \mathrm{U}(2)$, which is a symmetric space of type DIII (see \Cref{tab:symmetric_spaces}). 
The cosets consists of the action of $\mathrm{U}(2)$ within $\SO(4)$. A basis for $\g=\so(4)$ in terms of Pauli matrices is 
\begin{align*}
        \g &= i\:\Span\{IY, YI, XY, YX, YZ, ZY\},
    \end{align*}
which are all skew-symmetric. The subalgebra $\kk$ consists of all elements $A\in\g$ such that $J_2AJ_2=-A$, where $J_2=iYI$ as defined in \Cref{eq:matrices}. We find
\begin{align*}
    \kk=i\:\Span\{IY, YI, YX, YZ\} \cong \uu(2) 
\end{align*}
and
\begin{align*}
    \mm &= i\:\Span\{XY,ZY \}.
\end{align*}
The commutant and center of $\kk$ are given by $\g^\kk = \zz(\kk) = i\:\Span\{YI\}$.


\subsection{Stabilizing Gates}\label{app:stab}
\subsubsection{\texorpdfstring{$\SU(N)/\SU(N-1):$}{} Generalized direct state transformation \label{app:projective_1}}

\begin{figure}[htb!]
    \centering
    \includegraphics[width=0.65\textwidth]{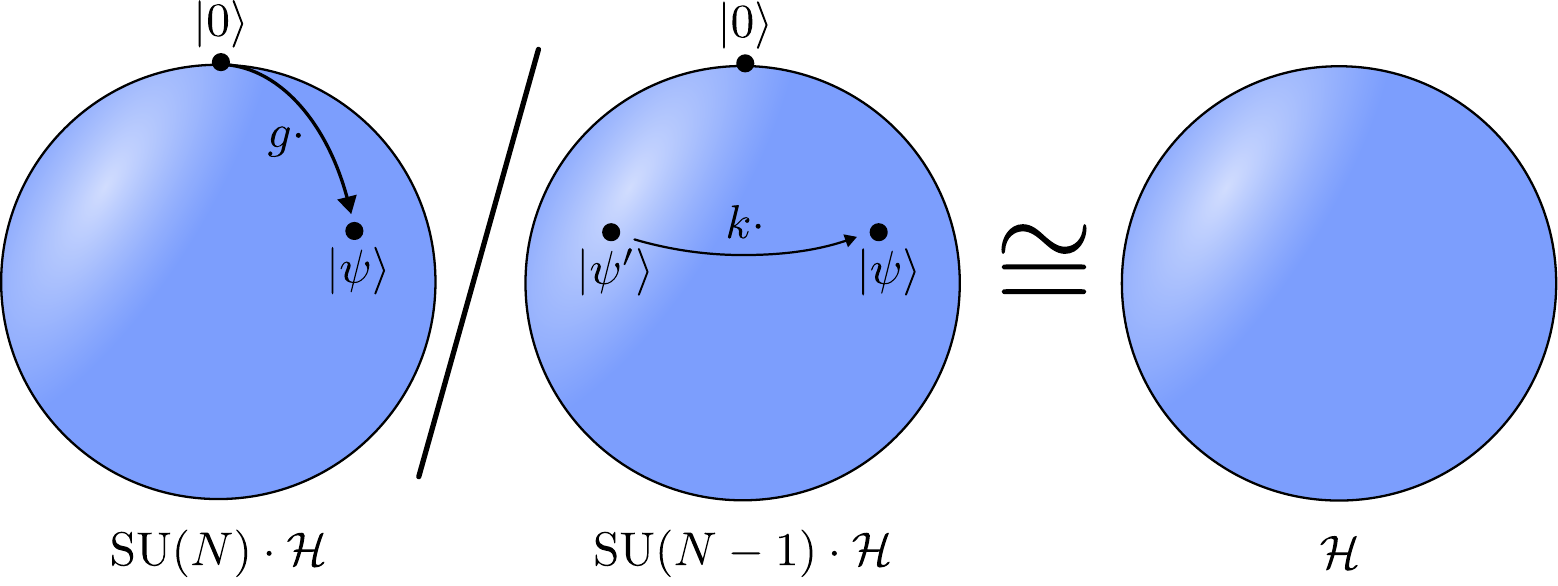}
    \hspace{10mm}
    \caption{\textbf{Schematic representation of the identification of two homogeneous spaces to a separate differentiable manifold.} Each $g\in\SU(N)$ can move any $\ket{\psi}\in\mathcal{H}$ to any other state $\ket{\psi'}\in\mathcal{H}$. The action $k\in \SU(N-1)$ acts on states in $\mathcal{H}$ while fixing $\ket{0}$. The quotient of the two groups thus allows us to uniquely identify each state in $\mathcal{H}$ with an action of $g \in \SU(N)$.
    }
    \label{fig:spheres_proj}
\end{figure}

Note that the action of $\SU(N)$ on the set of states in $\mathcal{H}=\mathbb{C}^N$ with unit norm is transitive: for every two such states $\ket{\psi}, \ket{\psi'}\in\mathcal{H}$, there exists a $g \in \SU(N)$ such that $g \cdot \ket{\psi} = \ket{\psi'}$. We consider the state $\ket{0}$. It is clear that this state is stabilized by $G_{\ket{0}} = \SU(N-1) \hookrightarrow \SU(N)$, which we can understand as the set of block diagonal matrices
\begin{align*}
    G_{\ket{0}} = \left(\begin{array}{@{}c|c@{}}
        1& 0 \\\hline
        0 & \begin{matrix}
             &  & \\
             & \SU(N-1) & \\
             &  & 
        \end{matrix}  
    \end{array}\right).
\end{align*}
The homogeneous space $G/G_{\ket{0}} = \SU(N)/\SU(N-1)$ therefore consists of the actions of $\SU(N)$ that move $\ket{0}$ to some other state $\ket{\psi}$ (see~\Cref{fig:spheres_proj}). We can identify each $\ket{\psi}\in\mathcal{H}$ with an action of $g$ on $\ket{0}$. Hence, we see that $G/G_{\ket{0}}$ is isomorphic to 
the unit sphere $S^{2N-1}$ in $\mathcal{H}$.
\subsubsection{\texorpdfstring{$\SU(N)/\SU(N-2):$}{} A fixed set of vectors}
A simple extension of this idea is the case where we consider the stabilizer of a set of states in $\mathcal{H}$, 
in which case the states are fixed pointwise.
For example, consider the set $\{\ket{0},\ket{1}\}$. Then its stabilizer consists of matrices of the form
\begin{align*}
    G_{\{\ket{0},\ket{1}\}} = G_{\ket{0}} \cap G_{\ket{1}} = \left(\begin{array}{@{}c|c@{}}
        \begin{array}{cc}
            1 & 0 \\
            0 & 1
        \end{array} & 0 \\\hline
        0 & \begin{matrix}
             &  & \\
             & \SU(N-2) & \\
             &  & 
        \end{matrix}  
    \end{array}\right).
\end{align*}

\subsubsection{\texorpdfstring{$\SU(N)/\U(N-1):$}{} Projective Hilbert spaces \label{app:projective_2}}

If instead of stabilizing $\ket{0}$, we want the coset $[0]$ given by the equivalence relation $\ket{0}\sim\lambda \ket{0}$ for $\lambda\in\mathbb{C}$ and $\lambda\neq 0$, we obtain the homogeneous symmetric space $\SU(N)/\U(N-1)$, which is the projective Hilbert space $\mathbb{C}\mathrm{P}^{N-1}$.

\begin{figure}[htb!]
    \centering
    \includegraphics[width=0.65\textwidth]{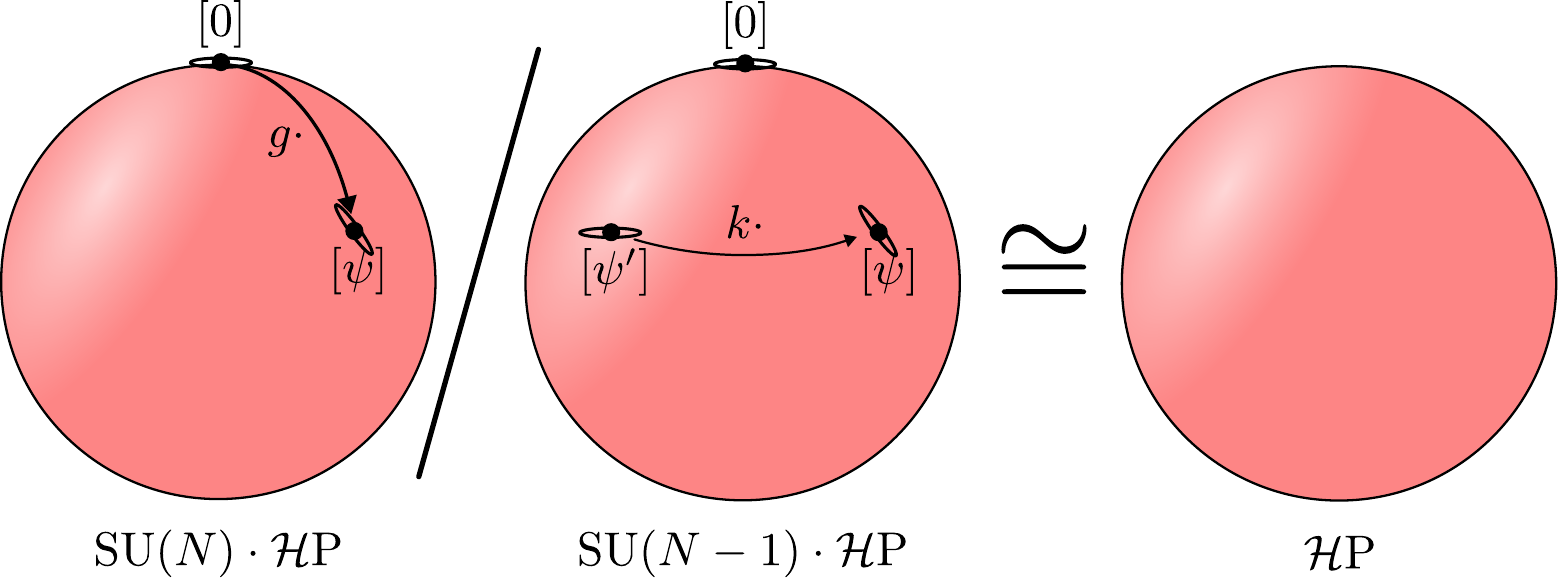}
    \hspace{10mm}
    \caption{\textbf{Schematic representation of the identification of two homogeneous spaces to a separate differentiable manifold.} The same argument as in \Cref{fig:spheres_proj} applies here, except that the states $\ket{\psi}\in\mathcal{H}$ are replaced with cosets in $\mathcal{H}\mathrm{P}$.}
    \label{fig:spheres_proj_2}
\end{figure}

Indeed, recall that $\mathcal{H} \mathrm{P} \equiv \mathbb{C}\mathrm{P}^{N-1}$ is the space of cosets $[\psi]$ under the equivalence relation $\ket{\psi}\sim \lambda \ket{\psi}$. It is straightforward to see that $[0]$ is stabilized by $\mathrm{S}(\U(1)\times \U(N-1)) \cong \U(N-1)$, which can be represented as matrices of the form
\begin{align*}
    G_{[0]} = \left(\begin{array}{@{}c|c@{}}
        e^{i\phi} & 0 \\\hline
        0 & \begin{matrix}
             &  & \\
             & B & \\
             &  & 
        \end{matrix}  
    \end{array}\right),\quad \phi\in\mathbb{R}, \; B\in\U(N-1), \; \det(B)=e^{-i\phi},\label{eq:stab_su_n_1}
\end{align*}
since $e^{i\phi}[0] = [0]$. The homogeneous space $\SU(N)/\U(N-1)$ is then the space of actions of $\SU(N)$ that move $[0]$ to another coset $[\psi]$ (see~\Cref{fig:spheres_proj_2}). 

In addition to being a homogeneous space, $\SU(N)/\U(N-1)$ is also a symmetric space
 of type AIII with $p=1$, $q=N-1$ (see \Cref{tab:symmetric_spaces}).
In fact, the above construction is one of the three cases of $\SO(N)/\mathrm{O}(N-1)$, $\SU(N)/\U(N-1)$ and $\Sp(N)/(\Sp(1) \times \Sp(N-1))$, which correspond to the real, complex and quaternionic projective spaces $\mathbb{R}\mathrm{P}^{N-1}$, $\mathbb{C}\mathrm{P}^{N-1}$ and $\mathbb{H}\mathrm{P}^{N-1}$, respectively. 
\clearpage

\subsubsection{\texorpdfstring{$\SU(p+q)/\mathrm{S}(\mathrm{U}(p)\times \mathrm{U}(q)):$ Complex Grassmannians}{}}\label{app:grassmannian}

The analysis for the projective spaces can be extended even further to the cases where the cosets are under some non-Abelian symmetry. For example, let 
$S$ be a $p$-dimensional subspace of $\mathbb{C}^N$ (for a fixed $1\le p<N$).
Then the stabilizer $G_S$ of $S$ in $G=\SU(N)$ can be identified with all matrices of the form
\begin{align*}
    \left(\begin{array}{@{}c|c@{}}
        A& 0 \\\hline
        0 & \begin{matrix}
             &  & \\
             & B & \\
             &  & 
        \end{matrix}  
    \end{array}\right),
    \qquad A\in \U(p), \;\; B\in \U(q), \;\; \det(A)\det(B)=1,
\end{align*}
where $N = p + q$. Hence, $K=G_S = \mathrm{S}(\mathrm{U}(p)\times \mathrm{U}(q))$.
The homogeneous space $G/G_{[\psi]}$ is called a complex Grassmannian $\mathrm{Gr}_p(\mathbb{C}^N)$;
this is the space of all $p$-dimensional subspaces of $\mathbb{C}^N$.
This is a symmetric space of type AIII (see \Cref{tab:symmetric_spaces}).
Analogously to the projective spaces, there are real, complex and quaternionic Grassmannian symmetric spaces, which are given by $\SO(p+q)/\mathrm{SO}(p)\times \mathrm{SO}(q)$, $\SU(p+q)/\mathrm{S}(\mathrm{U}(p)\times \mathrm{U}(q))$ and $\Sp(p+q)/\Sp(p)\times \Sp(q)$, respectively.

An example is the case $G = \SU(8)$ and $K = \mathrm{S}(\U(2)\times\U(6))$. Here, 
\begin{align*}
    \kk &= i\,\Span\{IIX, ZIX,IIZ,ZIZ ,
-IXX + ZXX ,-IXZ + ZXZ ,-IYX + ZYX ,-IYZ + ZYZ ,\\
&IZX, ZZX, IZZ, ZZZ,
-XIX + XZX ,-XIZ + XZZ,-XXX -YYX ,-XXZ -YYZ ,\\
& -XYX + YXX,-XYZ + YXZ,-XIY + XZY,-XXI -YYI ,
-YIX + YZX,-YIZ + YZZ,\\
& -XYY + YXY, XII -XZI, XXY + YYY, XYI -YXI,-YIY + YZY, -YII + YZI ,
IIY, ZIY ,\\
&-IXI + ZXI ,-IXY + ZXY ,-IYI + ZYI ,-IYY + ZYY ,
IZI, ZZI, IZY, ZZY, ZII\},
\end{align*}
and
\begin{align*}
    \mm  &= i\,\Span\{
    -IXI -ZXI,-IXY -ZXY,-IYI -ZYI,-IYY -ZYY, -XII -XZI, -XIY -XZY, \\
   &-XXI + YYI, -XXY + YYY, -XYI -YXI,-XYY -YXY, XIX + XZX, XIZ + XZZ,\\
   & -YII -YZI, -YIY -YZY, XYX + YXX, XYZ + YXZ, -XXX + YYX, -XXZ + YYZ, \\ & YIX + YZX, YIZ + YZZ,IXX + ZXX, IXZ + ZXZ, IYX + ZYX, IYZ + ZYZ \}.
\end{align*}
The commutant $\g^\kk$ is trivial.

\subsubsection{\texorpdfstring{$\SO(4)/(1\times \SO(2) \times 1):$}{} Charge preserving gates}\label{ex:so4_I_so2}

Consider a two-qubit quantum system. A basis for the Hilbert space $(\mathbb{C}^2)^{\otimes 2}$ is given by $\{\ket{00},\ket{01},\ket{10},\ket{11}\}$. We can label the states by their occupancy; $\ket{00} \to \ket{\Bar 0}$, $\{\ket{01},\ket{10}\}\to \ket{\Bar1}$ and $\ket{11}\to \ket{\Bar2}$. The subspace spanned by $\ket{01}$ and $\ket{10}$ over the real numbers defines a coset $[\bar 1]$ determined by the equivalence relation $g\cdot \ket{\Bar1} \sim \ket{\Bar1}$ with $g\in\SO(2)$. We see that the stabilizer of $[\bar 1]$ is

\begin{align*}
    G_{[\bar 1]} = \begin{pmatrix}
        1 & 0 & 0\\
        0 & \SO(2) & 0 \\
        0 & 0 & 1
    \end{pmatrix} \cong \SO(2).
\end{align*}
In the fermionic quantum computing literature, the elements of the above stabilizer are known as Givens rotations. Such an operation can rotate freely between the $\ket{01}$ and $\ket{10}$ quantum states, while keeping the $\ket{00}$ and $\ket{11}$ states fixed. Such operations are useful when one is interested in preserving the total occupation number. We can construct the homogeneous space $G/G_{[\bar 1]}$, for which we find $\kk  = i\:\Span\{0\oplus Y\oplus0\}$ and
\begin{align*}
    \mm  = i\:\Span\{&YI,YZ, ZY,\sqrt{2}IY+ XY+YX,-\sqrt{2}IY+ XY+YX \}.
\end{align*}
A choice of a maximal Abelian subalgebra is $\hh = i\:\Span\{XY, YX\}$. We find the commutant to be $\g^\kk=i\:\Span\{XY-YX, XY+YX\}$, and $\zz(\kk) = i\:\Span\{XY-YX\} = \kk$. Note that $XY-YX = -2(0\oplus Y\oplus0)$.
\end{document}